\documentclass[11pt,aps,showpacs,notitlepage,nofootinbib]{revtex4-1}
\usepackage{amssymb,amsmath}
\usepackage{bm}
\usepackage[pdftex]{graphicx}
\usepackage{subfig}
\usepackage{color}
\usepackage{amsthm}
\usepackage[update,prepend]{epstopdf}

\usepackage{footnote}

\newcommand{\beq}{\begin{equation}}
\newcommand{\beqn}{\begin{eqnarray}}
\newcommand{\eeq}{\end{equation}}
\newcommand{\eeqn}{\end{eqnarray}}

\begin{document}

\title{Preheating after multifield inflation with nonminimal couplings, III: \\ Dynamical Spacetime Results }
\author{Matthew P. DeCross$^1$\footnote{Now at the Department of Physics, University of Pennsylvania}, David I. Kaiser$^1$, Anirudh Prabhu$^1$\footnote{Now at the Department of Physics, Stanford University}, Chanda Prescod-Weinstein$^2$, and Evangelos I. Sfakianakis$^3$\footnote{Now at NIKHEF and Leiden University.} }
\email{\baselineskip 11pt Email addresses: mdecross@sas.upenn.edu ; dikaiser@mit.edu ; aniprabhu@stanford.edu ; cprescod@uw.edu ; \\ evans@nikhef.nl}
\affiliation{$^1$Department of Physics, 
Massachusetts Institute of Technology, Cambridge, Massachusetts 02139 USA
\\
$^2$ Department of Physics, University of Washington, Seattle, Washington 98195-1560
\\
$^3$Department of Physics, University of Illinois at Urbana-Champaign, Urbana, Illinois 61801}
\date{\today}
\begin{abstract}
This paper concludes our semi-analytic study of preheating in inflationary models comprised of multiple scalar fields coupled nonminimally to gravity. Using the covariant framework of Ref.~\cite{MultiPreheat1}, we extend the rigid-spacetime results of Ref.~\cite{MultiPreheat2} by considering both the expansion of the universe during preheating, as well as the effect of the coupled metric perturbations on particle production. The adiabatic and isocurvature perturbations are governed by different effective masses that scale differently with the nonminimal couplings and evolve differently in time. The effective mass for the adiabatic modes is dominated by contributions from the coupled metric perturbations immediately after inflation. The metric perturbations contribute an oscillating tachyonic term that enhances an early period of significant particle production for the adiabatic modes, which ceases on a time-scale governed by the nonminimal couplings $\xi_I$. The effective mass of the isocurvature perturbations, on the other hand, is dominated by contributions from the fields' potential and from the curvature of the field-space manifold (in the Einstein frame), the balance between which shifts on a time-scale governed by $\xi_I$. As in Refs.~\cite{MultiPreheat1,MultiPreheat2}, we identify distinct behavior depending on whether the nonminimal couplings are small ($\xi_I \lesssim {\cal O} (1)$), intermediate ($\xi_I \sim {\cal O} (1 - 10)$), or large ($\xi_I \geq 100$).
\end{abstract}
\pacs{98.80.Cq ; 95.30.Cq.  Preprint MIT-CTP/4849.}
\maketitle

\section{Introduction}

This paper continues the work of Refs.~\cite{MultiPreheat1,MultiPreheat2} by considering the early stage of post-inflation reheating in models that involve multiple scalar fields, each with a nonminimal coupling to gravity. Reheating is a critical epoch in cosmic history, connecting early-universe inflation with the successes of the standard big-bang scenario. (For reviews of post-inflation reheating, see Refs.~\cite{GuthKaiser,BTW,Allahverdi,Frolov,AHKK,MustafaBaumann}.) Both of the main features of the models we consider --- multiple fields \cite{LythRiotto,WandsReview,MazumdarRocher,MartinRingeval,VenninWands,GongMultifield} and nonminimal couplings \cite{Callan,Bunch,BirrellDavies,Buchbinder,ParkerToms,Odintsov1991,Markkanen2013} --- are well-motivated by high-energy theory, and encompass such models as Higgs inflation \cite{BezrukovShaposhnikov} (see also \cite{Futamase,Salopek,Fakir,Makino,DKconstraints,DKnGET}) and related models with attractor-like solutions \cite{KMS,GKS,KS,SSK,Lindealpha}. 

In Ref.~\cite{MultiPreheat1} we established a doubly-covariant formalism with which to study the behavior of field fluctuations and metric perturbations to linear order, which is gauge-invariant with respect to spacetime transformations ($x^\mu \rightarrow x^{\prime \mu}$) as well as invariant under field-space reparameterizations ($\phi^I \rightarrow \phi^{\prime I}$). We also demonstrated in Ref.~\cite{MultiPreheat1} that the strong single-field attractor that such models generically obey during inflation \cite{KMS,GKS,KS,SSK} persists through the early stages of reheating, for at least as long as the linearized approximation (in the field and metric perturbations) remains valid. The attractor makes these models safe from destabilization issues like the ones described in Ref.~\cite{SebastienRP}.

In Ref.~\cite{MultiPreheat2}, we applied the covariant formalism in the rigid-spacetime limit, in which we imagine holding the energy density fixed while sending $M_{\rm pl} \rightarrow \infty$. (The reduced Planck mass is given by $M_{\rm pl} \equiv 1/\sqrt{8\pi G} \simeq 2.43 \times 10^{18}$ GeV.) In that limit, we may neglect both the expansion of spacetime during reheating as well as the effects of the coupled metric perturbations \cite{AHKK}. Then the condensate(s) that had driven inflation oscillate periodically after the end of inflation, which can lead to efficient particle production via parametric resonance. Within the rigid-spacetime approximation, we studied the highly nonperturbative transfer of energy from the inflaton condensate into adiabatic and isocurvature modes, using the tools of Floquet theory. In the long-wavelength limit ($k \rightarrow 0$), we identified three distinct regimes, depending on whether the nonminimal coupling constants $\xi_I$ are small ($\xi_I \lesssim {\cal O} (1)$), intermediate ($\xi_\phi \sim {\cal O} (1 - 10)$), or large ($\xi_\phi \geq {\cal O} (100)$). For both adiabatic and isocurvature modes, the most efficient particle production occurs for strong couplings, $\xi_I \geq 100$, and approaches self-similar scaling solutions in the limit $\xi_I \rightarrow \infty$.

In this paper we consider the post-inflation dynamics while relaxing the assumption of a rigid spacetime, incorporating both the Hubble expansion and the contributions from the coupled metric perturbations. Consistent with earlier studies \cite{Taruya,BKM0,BKM,Sting,FinelliBrandenberger,ShinjiBassett,ChambersRajantie,BondFrolov,Bethke,Moghaddam}, we find that the coupled metric perturbations can have important effects on the preheating dynamics. In the models we consider here, the metric perturbations have a particularly strong effect on the behavior of the adiabatic modes, dominating their effective mass at early times with an oscillating, tachyonic contribution. Within the single-field attractor, on the other hand, the metric perturbations play a minimal role for the isocurvature modes. Instead, the behavior of the isocurvature modes is governed by the changing ratio of two other contributions to their effective mass: the term arising from (gradients of) the potential, and the term from the curved field-space manifold. We identify time-scales on which these distinct contributions dominate the dynamics, and how those time-scales vary with increasing $\xi_I$. As in our previous studies \cite{MultiPreheat1,MultiPreheat2}, we work to linear order in the perturbations, reserving for future study such nonlinear effects as backreaction of the created particles on the oscillating inflaton condensate \cite{AHKK,BKM,MustafaDuration,Figueroa}.

In Section~\ref{sec:Formalism} we briefly introduce our model and the covariant formalism with which to study linearized perturbations. Section~\ref{sec:Adiabatic} examines the behavior of the adiabatic modes for $k \sim 0$, and Section~\ref{sec:Isocurvature} examines the behavior of the isocurvature modes for $k \sim 0$. In Section~\ref{sec:Nonzerok}, we consider the contrasting behavior of adiabatic and isocurvature modes for nonzero wavenumber, studying the evolution of both sub-horizon and super-horizon modes. In Section~\ref{sec:Observational} we briefly consider possible observational consequences of the efficient preheating dynamics for this family of models, and in Section~\ref{sec:Higgs} we comment on potential implications of our analyses for reheating after Higgs inflation. Concluding remarks follow in Section~\ref{sec:Conclusions}. red In an Appendix, we consider the time-scale over which the background dynamics transition from a matter-dominated to radiation-dominated equation of state, and compare this cross-over time with the time-scales relevant for the amplification of adiabatic and isocurvature modes, as derived in Sections~\ref{sec:Adiabatic} and \ref{sec:Isocurvature}.

\section{Model and Formalism}
\label{sec:Formalism}

We consider models with $N$ real-valued scalar fields $\phi^I$ in $3+1$ spacetime dimensions. We use upper-case Latin letters to label field-space indices, $I, J = 1, 2, ..., N$; Greek letters to label spacetime indices, $\mu, \nu = 0, 1, 2, 3$; and lower-case Latin letters to label spatial indices, $i, j = 1, 2, 3$. The spacetime metric has signature $(-,+,+,+)$. The action in the Jordan frame is given by
\beq
S = \int d^4 x \> \sqrt{ - \tilde{g} } \left[ f (\phi^I ) \tilde{R} - \frac{1}{2} \delta_{IJ} \tilde{g}^{\mu\nu} \partial_\mu \phi^I \partial_\nu \phi^J - \tilde{V} (\phi^I ) \right] ,
\label{SJ}
\eeq
where tildes indicate Jordan-frame quantities. We perform a conformal transformation 
\beq
\tilde{g}_{\mu\nu} (x) \rightarrow g_{\mu\nu} (x) = \frac{2}{ M_{\rm pl}^2} f (\phi^I (x) ) \> \tilde{g}_{\mu\nu} (x) 
\label{conformal}
\eeq
to bring the action in the Einstein frame into the form \cite{DKconf,Abedi}
\beq
S = \int d^4 x \sqrt{-g} \left[ \frac{ M_{\rm pl}^2}{2} R - \frac{ 1}{2} {\cal G}_{IJ} (\phi^K ) g^{\mu\nu} \partial_\mu \phi^I \partial_\nu \phi^J - V (\phi^I ) \right] .
\label{SE}
\eeq
In the Einstein frame, the (curved) field-space manifold acquires a metric given by 
\beq
{\cal G}_{IJ} (\phi^K) = \frac{ M_{\rm pl}^2}{2 f (\phi^K )} \left[ \delta_{IJ} + \frac{ 3}{f (\phi^K )} f_{, I} f_{, J} \right] ,
\label{GIJ}
\eeq
where $f_{, I} = \partial f / \partial \phi^I$. (Explicit expressions for the components of ${\cal G}_{IJ}$ for our two-field model may be found in Appendix A of Ref.~\cite{MultiPreheat1}.) The potential in the Einstein frame is likewise stretched by the conformal factor:
\beq
V (\phi^I) = \frac{M_{\rm pl}^4}{4 f^2 (\phi^I ) } \tilde{V} (\phi^I) .
\label{VE1}
\eeq

We expand the scalar fields and spacetime metric to first order in perturbations. Because we are interested in the dynamics at the end of inflation, we consider scalar metric perturbations around a spatially flat Friedmann-Lema\^{i}tre-Robertson-Walker (FLRW) line element,
\beqn
\begin{split}
ds^2 &= g_{\mu\nu} (x) \> dx^\mu dx^\nu \\
&= - (1 + 2 A) dt^2 + 2 a \left( \partial_i B \right)  dx^i dt + a^2 \left[ ( 1 - 2 \psi ) \delta_{ij} + 2 \partial_i \partial_j E \right] dx^i dx^j ,
\label{ds}
\end{split}
\eeqn
where $a(t)$ is the scale factor. We also expand the fields,
\beq
\phi^I (x^\mu) = \varphi^I (t) + \delta \phi^I (x^\mu) . 
\label{phivarphi}
\eeq
To first order in the perturbations, we may then construct generalizations of the gauge-invariant Mukhanov-Sasaki variable (see Ref.~\cite{MultiPreheat1} and references therein):
\beq
Q^I = \delta \phi^I + \frac{\dot{\varphi}^I}{H} \psi .
\label{Qdef}
\eeq

To background order, the dynamics are governed by the coupled equations,
\beq
{\cal D}_t \dot{\varphi}^I + 3 H \dot{\varphi}^I + {\cal G}^{IJ} V_{, J} = 0
\label{eomvarphi}
\eeq
and
\beqn
\begin{split}
H^2 &= \frac{1}{3 M_{\rm pl}^2} \left[ \frac{1}{2} {\cal G}_{IJ} \dot{\varphi}^I \dot{\varphi}^J + V (\varphi^I ) \right] , \\
\dot{H} &= - \frac{1}{ 2 M_{\rm pl}^2 } {\cal G}_{IJ} \dot{\varphi}^I \dot{\varphi}^J ,
\end{split}
\label{Friedmann}
\eeqn
where overdots denote derivatives with respect to $t$, and the Hubble parameter is given by $H (t) = \dot{a} / a$. Covariant derivatives with respect to the field-space metric are given by ${\cal D}_J A^I = \partial_J A^I + \Gamma^I_{\> JK} A^K$ for a field-space vector $A^I$, from which we may construct the (covariant) directional derivative with respect to cosmic time, ${\cal D}_t A^I = \dot{\varphi}^J {\cal D}_J A^I = \dot{A}^I + \Gamma^I_{\> JK} \dot{\varphi}^J A^K$, where the Christoffel symbols $\Gamma^I_{\> JK} (\varphi^L)$ are constructed from ${\cal G}_{IJ} (\varphi^K)$. The gauge-invariant perturbations obey
\beq
{\cal D}_t^2 Q^I + 3 H {\cal D}_t Q^I + \left[ \frac{ k^2}{a^2} \delta^I_{\> J} + {\cal M}^I_{\> J} \right] Q^J = 0 ,
\label{eomQ}
\eeq
where the mass-squared matrix is given by
\beq
{\cal M}^I_{\> J} \equiv {\cal G}^{IK} \left( {\cal D}_J {\cal D}_K V \right) - {\cal R}^I_{\> LMJ} \dot{\varphi}^L \dot{\varphi}^M - \frac{1}{M_{\rm pl}^2 a^3} {\cal D}_t \left( \frac{ a^3}{H} \dot{\varphi}^I \dot{\varphi}_J \right)
\label{MIJdef}
\eeq
and ${\cal R}^I_{\> LMJ}$ is the Riemann tensor constructed from ${\cal G}_{IJ} (\varphi^K)$. The term in Eq.~(\ref{MIJdef}) proportional to $1 / M_{\rm pl}^2$ arises from the coupled metric perturbations.

As in Refs.~\cite{MultiPreheat1,MultiPreheat2}, we consider a two-field model, $\phi^I = \{\phi, \chi \}^T$, with nonminimal couplings (in the Jordan frame) given by
\beq
f (\phi^I ) = \frac{1}{2} \left[ M_{\rm pl}^2 + \xi_\phi \phi^2 + \xi_\chi \chi^2 \right]
\label{f}
\eeq
and Jordan-frame potential
\beq
\tilde{V} (\phi^I ) = \frac{ \lambda_\phi}{4} \phi^4 + \frac{ g}{2} \phi^2 \chi^2 + \frac{ \lambda_\chi}{4} \chi^4 .
\label{VJ}
\eeq
The Einstein-frame potential $V (\phi^I)$ develops ridges and valleys for the generic case in which $\xi_\phi \neq \xi_\chi$ and/or $\lambda_\phi \neq g \neq \lambda_\chi$, and the background dynamics exhibit strong single-field attractor behavior \cite{MultiPreheat1,KMS,GKS,KS,SSK}. Given our covariant framework, we may always perform a field-space rotation such that the attractor lies along the direction $\chi = 0$ in field space, without loss of generality. Within such an attractor, the field-space metric becomes effectively diagonal, with ${\cal G}_{\phi \chi} = {\cal O} (\chi ) \sim 0$ \cite{MultiPreheat1}.

We rescale the perturbations, $Q^I (x^\mu) \rightarrow X^I (x^\mu) / a(t)$, and quantize, $X^I \rightarrow \hat{X}^I$, expanding $\hat{X}^\phi$ and $\hat{X}^\chi$ in sets of creation and annihilation operators and associated mode functions. Whereas in Ref.~\cite{MultiPreheat1} we worked in terms of conformal time, $d \eta = dt / a(t)$, in this paper we will use cosmic time, $t$, since our numerical routines are efficient at evolving $\varphi^I (t), H(t)$, and $X^I (t)$ in terms of $t$. Within the single-field attractor, the mode functions effectively decouple \cite{MultiPreheat1}:
\beqn
\begin{split}
\ddot{v}_k &+ H \dot{v}_k + \left( \frac{ k^2}{a^2} + m_{\rm eff, \phi}^2 (t) \right) v_k \simeq 0 , \\
\ddot{z}_k &+ H \dot{z}_k + \left( \frac{k^2}{a^2} + m_{\rm eff,\chi}^2 (t) \right) z_k \simeq 0 .
\end{split}
\label{vzeom}
\eeqn
The Hubble-drag term enters as $H \dot{v}_k$ (rather than $3H \dot{v}_k$, as in Eq.~\eqref{eomQ}), because we have scaled the fluctuations, $Q^I = X^I / a \propto a^{-1} \{ v_k, z_k \}^T$. Within the attractor (along $\chi = 0$), $v_k$ is the mode function for perturbations along the adiabatic direction and $z_k$ is the mode function for perturbations along the isocurvature direction. Their effective masses consist of four distinct contributions \cite{MultiPreheat1}:
\beq
m_{ {\rm eff}, I}^2 = m_{1, I}^2 + m_{2, I}^2 + m_{3,I}^2 + m_{4,I}^2 ,
\label{meffgeneral}
\eeq
with
\beqn
\begin{split}
m_{1, \phi}^2 &\equiv {\cal G}^{\phi K} \left( {\cal D}_\phi {\cal D}_K V \right) , \\
m_{2, \phi}^2 &\equiv - {\cal R}^\phi_{\> LM \phi} \dot{\varphi}^L \dot{\varphi}^M , \\
m_{3, \phi}^2 &\equiv - \frac{1}{ M_{\rm pl}^2 a^3} \delta^\phi_{\> I} \delta^J _{\> \phi} \> {\cal D}_t \left( \frac{a^3}{H} \dot{\varphi}^I \dot{\varphi}_J \right) , \\
m_{4, \phi}^2 &\equiv - \frac{1}{6} R ,
\end{split}
\label{miphi}
\eeqn
where $R$ is the spacetime Ricci curvature scalar; comparable expressions follow for the contributions to $m_{\rm eff , \chi}^2$. We note that $m_{1, I}^2$ arises from (covariant gradients of) the potential, $m_{2, I}^2$ from the curvature of the field-space manifold, $m_{3, I}^2$ from the coupled metric perturbations, and $m_{4, I}^2$ from the expanding spacetime. The form of each of these contributions differs from the case of minimally coupled fields. The term $m_{2, I}^2$, in particular, has no analogue for models with minimally coupled fields and canonical kinetic terms, and can play important roles in the dynamics both during and after inflation \cite{MultiPreheat1,MultiPreheat2,KMS,GKS,KS,SSK,SebastienRP}. Meanwhile, within the single-field attractor along $\chi = 0$, the energy densities for adiabatic and isocurvature perturbations take the form \cite{MultiPreheat1}
\beqn
\begin{split}
\rho_k^{(\phi)} &= \frac{1}{2} \left[ \vert \dot{v}_k \vert^2 + \left( \frac{ k^2}{a^2} + m_{\rm eff,\phi}^2 \right) \vert v_k \vert^2 \right] + {\cal O} (\chi^2 ) , \\
\rho_k^{(\chi)} &= \frac{1}{2} \left[ \vert \dot{z}_k \vert^2 + \left( \frac{ k^2}{a^2} + m_{\rm eff,\chi}^2 \right) \vert z_k \vert^2 \right] + {\cal O} (\chi^2 ) ,
\end{split}
\label{rhodef}
\eeqn
again keeping in mind that $Q^\phi \sim v_k / a(t)$ and $Q^\chi \sim z_k / a(t)$. We measure particle production with respect to the instantaneous adiabatic vacuum, $\vert 0 (t_{\rm end}) \rangle$, which minimizes the energy densities $\rho_k^{(I)}$ at the end of inflation \cite{AHKK}.

For evolution within an attractor along the direction $\chi = 0$, the Hubble scale at the end of inflation satisfies 
\beq
H (t_{\rm end} ) \simeq 0.4 \sqrt{ \frac{ \lambda_\phi}{ 12 \xi_\phi^2 } } \> M_{\rm pl},
\label{Hend}
\eeq
and the background field's frequency of oscillation at the end of inflation satisfies $\omega > H$ for all $\xi_\phi \geq 0$, with $\omega / H \simeq 1.5$ for $\xi_\phi \leq 0.1$ and $\omega / H \simeq 4$ for $\xi_\phi \geq 10$ \cite{MultiPreheat1}. In particular, both the period of oscillation $T = 2 \pi / \omega$ and the Hubble time $H^{-1}$ scale as $\xi_\phi$ for large $\xi_\phi$. Hence we may capture the relevant dynamics by rescaling our time variable $t \rightarrow \tilde{t} \equiv \sqrt{ \lambda_\phi } \, M_{\rm pl} \,  t / \xi_\phi \propto H (t_{\rm end})\> t$. For the remainder of this paper, we will set $\tilde{t}_{\rm end} = 0$, where $\tilde{t}_{\rm end}$ is the time at which inflation ends and preheating begins.

\begin{figure}
\centering
\includegraphics[width=.48\textwidth]{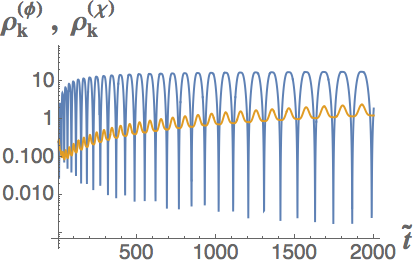} \quad \includegraphics[width=.48\textwidth]{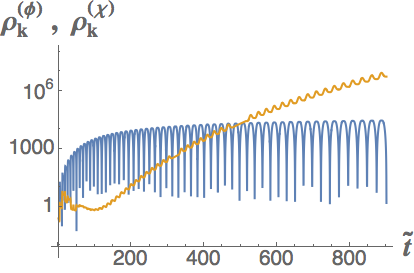} 
\\
\includegraphics[width=.48\textwidth]{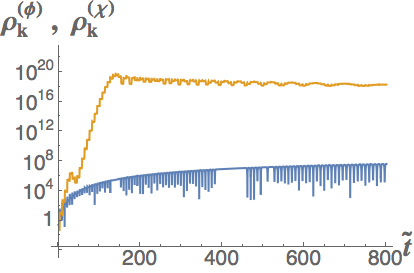} \quad \includegraphics[width=.48\textwidth]{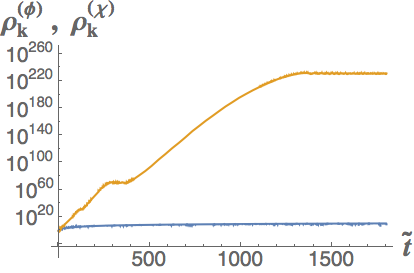} 
  \caption{\small \baselineskip 11pt
The evolution of the energy densities $\rho^{(\phi)}_k$ (blue) and $\rho^{(\chi)}_k$ (gold) as functions of $\tilde{t} = \sqrt{\lambda_\phi} \, M_{\rm pl} \, t / \xi_\phi$ for $k=0$. The parameter ratios are chosen as $\xi_\chi / \xi_\phi =0.8$, $g/\lambda_\phi=1$, and $\lambda_\chi / \lambda_\phi = 1.25$. ({\it Top, left to right}) $\xi_\phi=1,10$; ({\it bottom, left to right}) $\xi_\phi = 10^2,10^3$. }
\label{fig:allresultsintro}
\end{figure}

In Fig.~\ref{fig:allresultsintro} we show the growth of $\rho^{(\phi)}_k$ and $\rho^{(\chi)}_k$ for $k = 0$ as we increase the nonminimal couplings $\xi_I$. In particular, we select $\xi_\chi = 0.8 \> \xi_\phi$, $g = \lambda_\phi$, and $\lambda_\chi = 1.25 \> \lambda_\phi$, such that all of the relevant ratios among couplings are ${\cal O} (1)$, and hence no softly broken symmetry exists \cite{MultiPreheat2}. 
As in Refs.~\cite{MultiPreheat1,MultiPreheat2}, we identify three distinct regimes, governed by the magnitude of $\xi_I$: inefficient growth of either set of perturbations for $\xi_I \leq {\cal O} (1)$ in the limit $k \rightarrow 0$; modest growth of adiabatic modes for $\xi_I \sim {\cal O} (10)$; and rapid growth for both adiabatic and isocurvature modes for $\xi_I \geq {\cal O} (100)$. Moreover, the balance shifts between adiabatic and isocurvature modes between the intermediate- and large-$\xi_I$ regimes, with especially rapid growth of $\rho^{(\chi)}_k$ for $\xi_I \rightarrow \infty$. Hence we find that the growth within the broad-resonance regime that we analyzed in Ref.~\cite{MultiPreheat2} is robust, even when we relax the assumption of a rigid spacetime. In the next two sections, we analyze semi-analytically the behavior shown in Fig.~\ref{fig:allresultsintro}.

\section{Results: Adiabatic Modes}
\label{sec:Adiabatic}

Within the single-field attractor, there are two significant components of the effective mass $m_{\rm eff,\phi}^2$ for the adiabatic perturbations during preheating: $m_{1, \phi}^2$ from the potential and $m_{3, \phi}^2$ from the metric perturbations. Within the single-field attractor, $m_{2, \phi}^2 \sim {\cal O} (\chi \dot{\chi} ) \sim 0$, and $m_{4, \phi}^2 = - R / 6 \sim {\cal O} (1) \> H^2$ remains much smaller than the other terms \cite{MultiPreheat1}.

\begin{figure}
\centering
\includegraphics[width=0.45\textwidth]{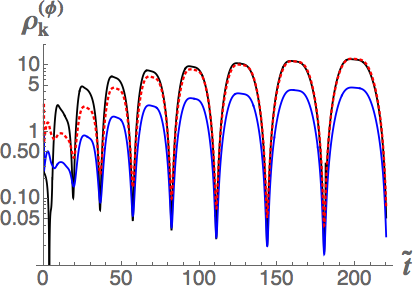}\quad \includegraphics[width=0.45\textwidth]{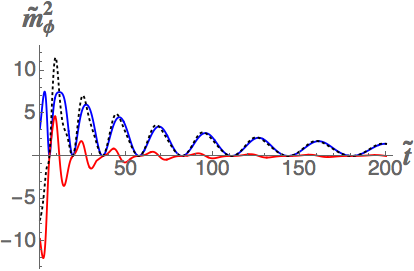}\\
\includegraphics[width=0.45\textwidth]{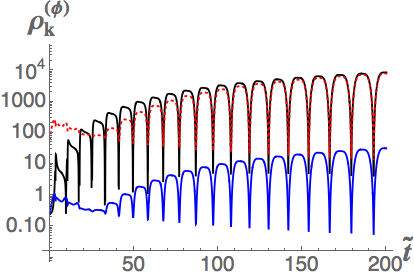}\quad \includegraphics[width=0.45\textwidth]{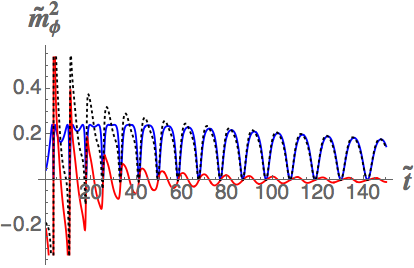}\\
\includegraphics[width=0.45\textwidth]{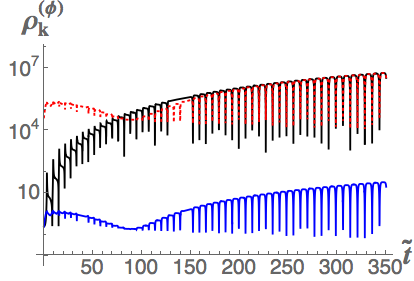}\quad \includegraphics[width=0.45\textwidth]{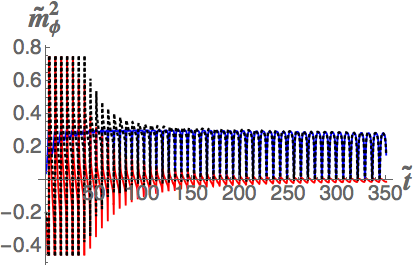}\\
\includegraphics[width=0.45\textwidth]{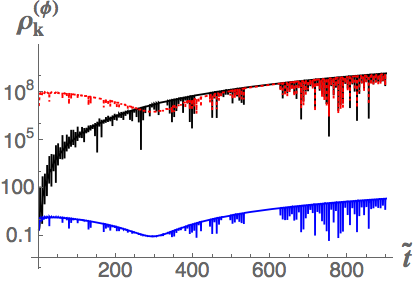}\quad \includegraphics[width=0.45\textwidth]{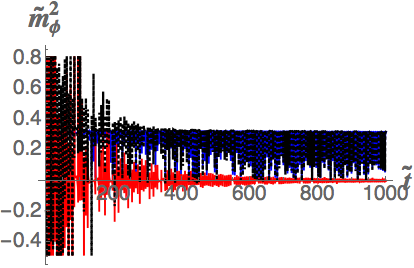}
\caption{\small \baselineskip 11pt ({\it Left}) The energy density for the adiabatic perturbations $\rho_k^{(\phi)}$ for $k = 0$ as a function of $\tilde{t} = \sqrt{\lambda_\phi} \, M_{\rm pl} \, t / \xi_\phi$, computed using the full effective mass (black) and  by neglecting the coupled metric perturbations (blue). The red dotted curve is a shifted version of the blue curve, to help indicate the time-scales on which $m_{1,\phi}^2$ or $m_{3,\phi}^2$ dominate $m_{\rm eff,\phi}^2$. ({\it Right}) The rescaled adiabatic effective mass $\tilde{m}_{\rm eff,\phi}^2 = \xi_\phi^2 m_{\rm eff,\phi}^2$ (back dotted) and its two main components, $\tilde{m}_{1,\phi}^2$ (blue) from the potential, and $\tilde{m}_{3, \phi}^2$ (red) from the metric perturbations. The nonminimal coupling is $\xi_\phi=1,10,10^2,10^3$ (top to bottom).}
\label{fig:adiabaticpanel}
\end{figure}

In order to distinguish the effects from $m_{1,\phi}^2$ and $m_{3, \phi}^2$ and examine their dependence on time and $\xi_I$, we compute $\rho_k^{(\phi)}$ first by neglecting $m_{3, \phi}^2$ and compare with the results computed using the full effective mass, $m_{\rm eff,\phi}^2$. The results for the $k = 0$ mode and different values of $\xi_\phi$ are shown in Fig.~\ref{fig:adiabaticpanel}, from which two main results become clear.
First, the contribution from the coupled metric perturbations, $m_{3, \phi}^2$, dominates during times immediately after the end of inflation, yielding a significant, oscillating tachyonic contribution to $m_{\rm eff,\phi}^2$ that drives significant particle production. The contribution from the potential, $m_{1,\phi}^2$, comes to dominate $m_{\rm eff,\phi}^2$ at later times, such that the energy density computed with either $m_{\rm eff,\phi}^2$ or $m_{1,\phi}^2$ alone share the same behavior after a characteristic time-scale.

We start by examining the time when the metric perturbations dominate the growth rate of the adiabatic modes, as a function of $\xi_\phi$, focusing on $\xi_\phi \ge10$. 
The maximum value (positive or negative) of $m_{3, \phi}^2$ may be well-fit numerically by 
\beq
{\rm max} (m^2_{3,\phi}) \simeq 2 \sqrt 6 \left | \ddot \phi_{\rm max} \right | ,
\eeq
where overdots denote derivatives with respect to $\tilde{t}$. We can see from the right panels of Fig.\ \ref{fig:adiabaticpanel} that the coupled metric perturbations become unimportant when ${\rm max}(m^2_{3,\phi}) \lesssim{1\over 3}{\rm max}(m^2_{1,\phi})$. Numerical evaluation of the relevant functions for several values of $\xi_\phi \ge 10$ pinpoints this crossover as occurring at $\tilde{t}_{\rm cross} \sim 9.5 \sqrt{\xi_\phi }$. Upon using Eq.~(\ref{Hend}) and the relation $\tilde{t} = \sqrt{\lambda_\phi} \, M_{\rm pl} \, t / \xi_\phi$, this is equivalent to $H_{\rm end} \> t_{\rm cross} \sim 1.1 \>\sqrt{\xi_\phi}$, where $H_{\rm end} \equiv H (t_{\rm end})$ is the value of the Hubble parameter at the end of inflation. The exact constant of proportionality is not particularly important; the point is that the growth of adiabatic modes for $\xi_\phi \geq 10$ is dominated by the effects of the coupled metric perturbations at early times, up to around $H_{\rm end} \> t_{\rm cross} \sim {\cal O} (1) \sqrt{\xi_\phi}$.

As we can see from the blue curves in the lefthand panels of Fig.\ \ref{fig:adiabaticpanel}, the times $t < t_{\rm cross}$ correspond to no particle production stemming from the potential term, $m_{1,\phi}^2$. Looking closely at the form of $m_{1, \phi}^2$ in the righthand panels of Fig.~\ref{fig:adiabaticpanel}, we see that during this early phase $m_{1, \phi}^2$ is qualitatively different than its late-time form. At early times, $m_{1,\phi}^2$ exhibits one global minimum at $m_{1, \phi}^2 = 0$ and one local minimum at $m_{1,\phi}^2 > 0$ for each period of oscillation. As long as the existence of the local minimum is pronounced, particle production due to $m_{1,\phi}^2$ is suppressed. However, the distance between the local minimum and the two neighboring maxima decreases as a function of time. Once the local minimum vanishes, the energy density in the adiabatic modes begins to grow as a power law, exactly matching the late-time growth rate of $\rho_k^{(\phi)}$ when calculated using the full effective mass, $m_{\rm eff,\phi}^2$. This late-time behavior matches what we found in Ref.~\cite{MultiPreheat2} using Floquet theory in the rigid-spacetime limit: the $k = 0$ mode for the adiabatic perturbations exhibited at most power-law growth, rather than exponential growth.

\begin{figure}
\includegraphics[width=0.48\textwidth]{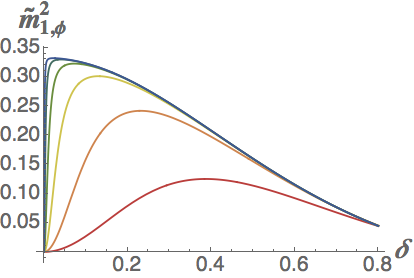}\quad \includegraphics[width=0.48\textwidth]{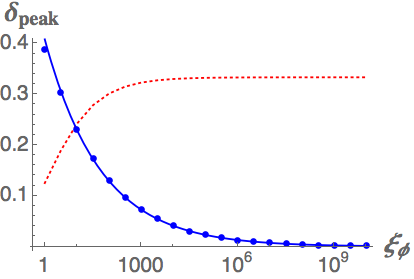}\\
\caption{\small \baselineskip 11pt ({\it Left}) The rescaled mass $\tilde{m}_{1,\phi}^2 = \xi_\phi^2 m_{1,\phi}^2$ as a function of $\delta = \sqrt{\xi_\phi} \> \phi / M_{\rm pl}$, for $\xi_\phi =1,10,10^2,10^3,10^4,10^5$ (color coded in a rainbow scale with red corresponding to $\xi_\phi=1$). {(\it Right}) The maximum value of $\tilde{m}_{1,\phi}^2$ (red dotted) and the value of $\delta_{\rm peak}$ for which it occurs (blue dots) as a function of $\xi_\phi$, along with the fitting curve $\delta_{\rm peak} = 0.4 \> \xi_\phi^{-1/4}$ (blue solid).
}
\label{fig:deltamaxphi}
\end{figure}

We may evaluate the time at which the local minimum in $m_{1, \phi}^2$ vanishes. During each oscillation, the field passes through the interval $0 \leq \delta \leq \delta_{\rm max}$, where $\delta_{\rm max}$ is the amplitude of oscillation, and $\delta (t) \equiv \sqrt{\xi_\phi} \> \phi (t) / M_{\rm pl}$. We label $\delta_{\rm peak}$ as the value of $\delta$ at which $m_{1, \phi}^2$ has a maximum; $\delta_{\rm peak}$ depends on $\xi_\phi$, as shown in Fig.~\ref{fig:deltamaxphi}. As long as $\delta_{\rm max} \geq \delta_{\rm peak}$, the term $m_{1,\phi}^2$ will oscillate between the global minimum at zero, the local maxima, and the local minimum. Once $\delta_{\rm max}$ becomes less than $\delta_{\rm peak}$, however, $m_{1,\phi}^2$ will only oscillate between zero and $m_{1,\phi}^2 (\delta_{\rm max} )$, exhibiting just one minimum and one maximum for each period of the background field's oscillation. 

As shown on the righthand side of Fig.~\ref{fig:deltamaxphi}, the quantity $\delta_{\rm peak}$ is well fit numerically by the curve $\delta_{\rm peak} \simeq 0.4 \> \xi_{\phi}^{-1/4}$ across nine orders of magnitude. The amplitude of oscillation, $\delta_{\rm max}$, will redshift due to Hubble friction, having begun at $\delta_{\rm max} (t_{\rm end}) = 0.8$ for $\xi_\phi \geq 1$ \cite{MultiPreheat1}; at early times after $t_{\rm end}$, the envelope of the background inflaton oscillations is well fit by the curve $\delta_{\rm max} \simeq 1.5 \> \tilde{t}^{-1/2}$ (as may be gleaned from Fig.~2 of Ref.~\cite{MultiPreheat1}). The local minimum of $m_{1,\phi}^2$ vanishes when $\delta_{\rm max} \leq \delta_{\rm peak} \simeq 0.4 \> \xi_\phi^{-1/4}$, which may be solved to give
\beq
H_{\rm end} \> t_{\rm peak} \simeq 1.7 \> \sqrt{\xi_\phi} \>. 
\label{tpeak}
\eeq
Our estimate of $t_{\rm peak}$ is quite close to the time $t_{\rm cross}$, when $m_{1,\phi}^2$ from the potential begins to dominate $m_{3,\phi}^2$ from the coupled metric perturbations. We therefore conclude that for $H_{\rm end} \> t \lesssim \sqrt{\xi_\phi}$, the metric perturbations dominate $m_{\rm eff,\phi}^2$, driving a tachyonic amplification of $\rho_k^{(\phi)}$, whereas for $H_{\rm end} \> t \gtrsim \sqrt{\xi_\phi}$, the potential term dominates and the adiabatic mode with $k = 0$ grows as a power-law rather than an exponential. The departure from the rigid-spacetime behavior found in Ref.~\cite{MultiPreheat2} grows with $\xi_\phi$, in the sense that for larger $\xi_\phi$, the metric perturbations drive an exponential amplification of $\rho_k^{(\phi)}$ in the long-wavelength limit for correspondingly longer (rescaled) times $\tilde{t}$.


\section{Results: Isocurvature Modes}
\label{sec:Isocurvature}

As expected, the phenomenology of isocurvature modes is richer than that of adiabatic modes. The case $\xi_\phi \sim {\cal O} (1 - 10)$ deserves special attention, since it exhibits opposite behavior from the large-$\xi_\phi$ regime. We see in Fig.\ \ref{fig:allresultsintro} that isocurvature fluctuations for $k=0$ and $\xi_\phi=10$ do not grow initially. However for $\xi_\phi \ge 100$, quasi-exponential growth begins right away. This is in keeping with the results we found in Refs.~\cite{MultiPreheat1,MultiPreheat2}, in which the isocurvature perturbations exhibited qualitatively different behavior in the intermediate- and large-$\xi_\phi$ regimes.

\begin{figure}
\centering
\includegraphics[width=.48\textwidth]{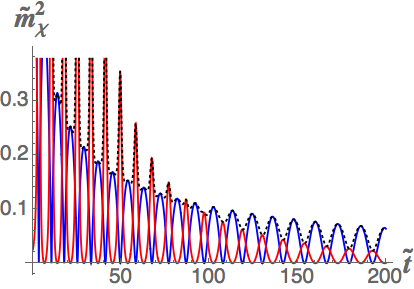} \quad \includegraphics[width=.48\textwidth]{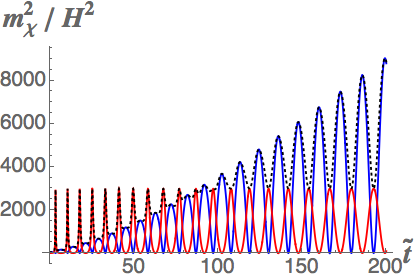}
  \caption{\small \baselineskip 11pt
  ({\it Left}) The rescaled effective mass for the isocurvature modes, $\tilde{m}_{\rm eff,\chi}^2 = \xi_\phi^2 m_{\rm eff,\chi}^2$ (black-dotted) along with the contributions from the potential, $\tilde{m}_{1,\chi}^2$ (blue), and from the curved field-space manifold, $\tilde{m}_{2,\chi}^2$ (red), for $\xi_\phi = 10$, as functions of $\tilde{t} = \sqrt{\lambda_\phi} \, M_{\rm pl} \, t / \xi_\phi$. ({\it Right}) The quantities $m_{\rm eff,\chi}^2$ (black-dotted), $m_{1,\chi}^2$ (blue), and $m_{2,\chi}^2$ (red) rescaled by the Hubble parameter. }
\label{fig:mchi_k10}
\end{figure}

Fig.\ \ref{fig:mchi_k10} shows the evolution of $m_{\rm eff,\chi}^2$ for $\xi_\phi=10$, along with its two main components within the single-field attractor, $m_{1,\chi}^2$ from the potential and $m_{2,\chi}^2$ from the curved field-space manifold. We see that $m_{2,\chi}^2$ dominates for early times and $m_{1,\chi}^2$ dominates at later times. Furthermore, the maximum amplitude of $m_{2,\chi}^2$ is proportional to $H^2 (t)$. This proportionality may be easily understood. Within the single-field attractor along $\chi = 0$ \cite{MultiPreheat1},
\beq
m^2_{2,\chi} = {1\over 2} {\cal R} \> {\cal G}_{\phi\phi} \dot \phi^2 \, ,
\label{m2chifirst}
\eeq
where ${\cal R}$ is the field-space Ricci scalar, not to be confused with the space-time Ricci scalar $R$. The peaks of $m^2_{2,\chi}$ occur when $\phi$ vanishes and $\dot \phi$ reaches its maxima. Using the expressions in Appendix A of Ref.~\cite{MultiPreheat1}, for large $\xi_I$ and $\phi=0$ we may easily derive
\beq
{\cal R}_{\rm max} \propto 6(1-\varepsilon) \xi_\phi^2 \> ,
\label{eq:calRmax}
\eeq
where $\varepsilon = (\xi_\phi - \xi_\chi ) / \xi_\phi$ is the ellipticity of the field-space potential. Furthermore, for motion along $\chi=0$ (within the single-field attractor) and with $\phi = 0$ (where $m_{2,\chi}^2$ peaks), we find ${\cal G}_{\phi\phi}=1$. Thus the peak of the contribution $m_{2,\chi}^2$ has a very simple expression:
\beq
{\rm max} \left ( m^2_{2,\chi} \right ) \propto  3(1-\varepsilon) \xi_\phi^2 \, \dot \phi^2 \>.
\eeq
After inflation the kinetic energy density (${\cal G}_{\phi\phi} \dot{\phi}^2 / 2$) and the potential energy density ($V (\phi,\chi)$) attain equal maximum values but oscillate out of phase. Averaged over a period of the inflaton oscillation, the Hubble parameter is
\beq
\langle H \rangle = {1\over \sqrt 6 M_{\rm pl}}  |\dot \phi_{\rm max}| \, ,
\eeq
where $|\dot \phi_{\rm max}|$ is the field's maximum velocity within one oscillation. Then one finds
\beq
{m^2_{2,\chi} \over \langle H \rangle^2} \propto (1-\varepsilon) \xi_\phi^2 \, ,
\label{m2Hcompare}
\eeq
indicating that the contribution to $m_{\rm eff,\chi}^2$ that arises from the curved field-space manifold, $m_{2,\chi}^2$, scales with $H^2 (t)$ after the end of inflation.

On the other hand, the contribution arising from the potential, $m_{1,\chi}^2$, grows compared to $H^2 (t)$, and thus dominates $m_{\rm eff,\chi}^2$ at late times. For $\xi_\phi = 10$, the terms $m_{1,\chi}^2$ and $m_{2,\chi}^2$ become approximately equal at $\tilde{t} \simeq 80$, while for $\tilde{t} \gtrsim 150$, the term $m_{1,\chi}^2$ clearly dominates. At these late times, $m_{\rm eff,\chi}^2$ oscillates quasi-sinusoidally and can drive parametric resonance, depending sensitively on wavenumber and couplings; this accounts for the late-time growth of $\rho_k^{(\chi)}$ for $\xi_\phi = 10$ in Fig.~\ref{fig:allresultsintro}.

The evolution of ${m^2_{1,\chi} / H^2}$ does not admit a simple solution, contrary to the adiabatic case ${m^2_{1,\phi} / H^2}$, for the following reason. In the case of $m^2_{1,\phi}$ the peak value occurrs at $\phi=0$, which drastically reduces the dependence of the peak on parameters, since all terms involving powers of $\phi$ vanish identically. In the case of  $m^2_{1,\chi}$, the peak occurs at some nonzero $\phi_{\rm max}$. Hence not only do higher powers of $\phi$ not vanish at the maximum of $m_{1,\chi}^2$, but, as $\phi (t)$ redshifts, terms such as $M_{\rm pl}^2 + \xi_\phi^2 \phi^2$ change their character over time. For $\xi_\phi \phi^2 / M_{\rm pl}^2 = {\cal O}(1)$, which is true immediately after inflation, $M_{\rm pl}^2 + \xi_\phi^2 \phi^2 \approx  \xi_\phi^2 \phi^2 $. However, as the amplitude of $\phi$ redshifts, the two terms in $M_{\rm pl}^2+ \xi_\phi^2 \phi^2$ become comparable.  Ultimately $M_{\rm pl}^2+ \xi_\phi^2 \phi^2 \to M_{\rm pl}^2$ as $t\to \infty $. Thus we can show that $m^2_{1,\chi} / H^2$ grows, but the nature of the growth does not admit a simple expression. 

\begin{figure}
\centering
\includegraphics[width=.48\textwidth]{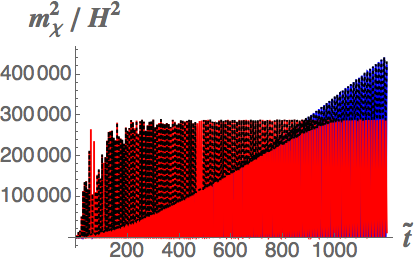} \quad \includegraphics[width=.48\textwidth]{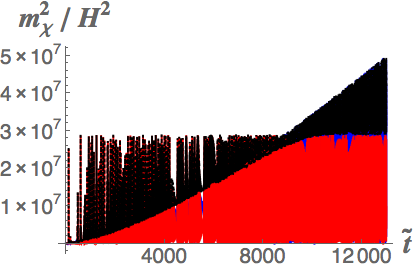}
  \caption{\small \baselineskip 11pt
The effective mass of the isocurvature perturbations, $m_{\rm eff,\chi}^2 / H^2$ (black-dotted), shown with the contributions from the potential, $m_{1,\chi}^2 / H^2$ (blue), and from the curved field-space manifold, $m_{2,\chi}^2 / H^2$ (red), for $\xi_\phi=10^2$ ({\it left}) and $\xi_\phi=10^3$ ({\it right}), as functions of $\tilde{t} = \sqrt{\lambda_\phi} \, M_{\rm pl} \, t / \xi_\phi$. }
\label{fig:mchi_k100_1000}
\end{figure}

In order to compare the two contributions $m^2_{1,\chi} $ and $m^2_{2,\chi} $, we may consider the former in two distinct regions:
\beq
 {m^2_{1,\chi} \over H^2}  
= \left\{ \begin{array}{ll}
         {\cal O}(\xi_\phi)  , & \delta = {\cal O}(1)  \\
          {\cal O}(\xi_\phi^2)   , & \delta = {\cal O}\left (\xi_\phi^{-1/2}  \right )  , \end{array}
         \right. 
         \label{m1Hcompare}
\eeq 
where $\delta(t)  = \sqrt{\xi_\phi} \> \phi  (t) / M_{\rm pl}$. We use $\delta = {\cal O}\left ( {1/ \sqrt{\xi_\phi} } \right ) $ as a point at which the amplitude of the background field has redshifted significantly, since for this field value the term $M_{\rm pl}^2 +\xi_\phi \phi^2$ changes its behavior.
We can immediately see that $m_{2,\chi}^2 / H^2 \propto \xi_\phi^2$ dominates initially, but will ultimately become subdominant. We may further analyze the $\xi_\phi$-dependence of the interplay between $m_{1,\chi}^2$ and $m_{2,\chi}^2$. The function $\delta (\tilde{t})$ asymptotes to one universal behavior in the limit $\xi_\phi \gg 1$ \cite{MultiPreheat1}. We noted above that the amplitude of the inflaton decays as $\delta \propto 1.5 \> \tilde{t}^{-1/2}$. This suggests that the crossover time between when $m_{2,\chi}^2$ and $m_{1,\chi}^2$ dominates should scale as $\tilde{t} \sim \xi_\phi$. We found that $m_{1,\chi}^2 \simeq m_{2,\chi}^2$ at $\tilde{t} \simeq 80$ for $\xi_\phi = 10$. Fig.~\ref{fig:mchi_k100_1000} shows that $m_{1,\chi}^2 \simeq m_{2,\chi}^2$ at $\tilde{t} \simeq 800$ for $\xi_\phi = 10^2$ and at $\tilde{t} \simeq 8000$ for $\xi_\phi = 10^3$, confirming that the crossover time scales as $\tilde{t} \sim \xi_\phi$.

\begin{figure}
\centering
\includegraphics[width=.48\textwidth]{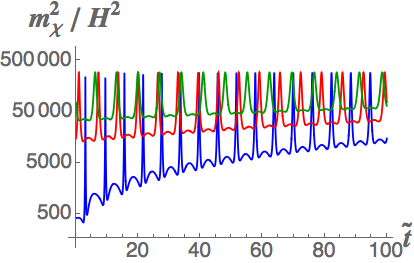}  \quad \includegraphics[width=.48\textwidth]{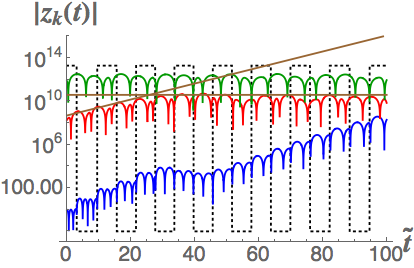}
  \caption{\small \baselineskip 11pt
({\it Left}) The effective mass for the isocurvature perturbations, $m_{\rm eff,\chi}^2 / H^2$, for ${\xi_\phi=100}$ as a function of $\tilde t = \sqrt{\lambda_\phi} \, M_{\rm pl} t / \xi_\phi$. The time interval of $0<\tilde t <300$ is folded into the interval $0<\tilde t <100$ as follows: $0<\tilde t <100$ (blue), $100< \tilde t <200$ (red), and $200 <\tilde t <300$ (green). ({\it Right}) The isocurvature mode $z_k(t)$ for $k = 0$, folded into the same time interval as the left panel. The green curve is shifted upwards for clarity. The black-dotted vertical lines correspond to the zero-crossings of the background field $\phi(t)$ in the interval $0<\tilde t<100$. The brown lines show the transition between exponential growth and slow red-shifting, which occurs around $\tilde{t} \sim 130$. }
\label{fig:mchi_k100}
\end{figure}

As we increase the nonminimal couplings from $\xi_I \sim {\cal O} (1 - 10)$ to $\xi_I \geq {\cal O} (100)$, we find a dramatic change in the behavior of the isocurvature modes in the long-wavelength limit. For $\xi_\phi = 100$, for example, $\rho_k^{(\chi)}$ exhibits strong initial growth (as shown in Fig.~\ref{fig:allresultsintro}), which ceases around $\tilde{t} \simeq 130$. As shown in Fig.~\ref{fig:mchi_k100}, $m_{\rm eff,\chi}^2$ is dominated at early times by $m_{2,\chi}^2$, with $m_{2, \chi}^2 / m_{1,\chi}^2 \simeq 100 = \xi_\phi$. (This comes from comparing the heights of the global maxima and the local maxima in $m_{\rm eff,\chi}^2$ at early times, and is in keeping with Eqs.~(\ref{m2Hcompare}) and (\ref{m1Hcompare}).) Using Figs.~\ref{fig:allresultsintro} and \ref{fig:mchi_k100}, we infer that the growth of $\rho_k^{(\chi)}$ ceases when the ratio falls to $m_{2,\chi}^2 / m_{1,\chi}^2 \simeq 15$, which occurs at around one-sixth the crossover time at which $m_{2,\chi}^2 = m_{1,\chi}^2$, corresponding to $\tilde{t} \sim 800 / 6 \sim 130$ for $\xi_\phi = 100$. Since the relevant transition times scale as $\tilde{t} \sim \xi_\phi$, we expect the period of rapid, initial growth in the case $\xi_\phi = 10^3$ to cease at $\tilde{t} \sim 8000 / 6 \sim 1300$, which indeed matches the behavior shown in Fig.~\ref{fig:allresultsintro}.

We may understand the transition out of the initial period of rapid growth for large $\xi_I$ as the change from the broad- to the narrow-resonance regime \cite{KLS}. At early times, the isocurvature mode $z_k (t)$ (with $k = 0$) oscillates multiple times for each oscillation of the inflaton field, as shown on the righthand side of Fig.~\ref{fig:mchi_k100}. These early oscillations drive a broad-resonance amplification during times when $m_{2,\chi}^2$ dominates $m_{\rm eff,\chi}^2$. At later times, however, the frequency of oscillation for $z_k (t)$ asymptotes toward the inflaton's oscillation frequency, which shifts the growth of $z_k (t)$ from broad resonance toward the less-efficient narrow-resonance regime. Although narrow-band resonances will drive exponential amplification in the rigid-spacetime limit, generically they become considerably less effective in an expanding universe \cite{KaiserPreh1,KLS}. Fig.~\ref{fig:mchi_k100}b indicates that $z_k (t)$ transitions from 2-3 oscillations per inflaton oscillation toward one oscillation per inflaton oscillation around $\tilde{t} \sim 130$, at the same time that the exponential growth of $z_k (t)$ ceases. 

\begin{figure}
\includegraphics[width=0.48\textwidth]{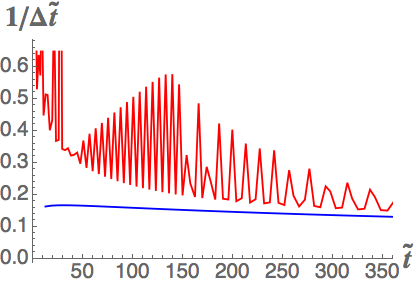}\quad \includegraphics[width=0.48\textwidth]{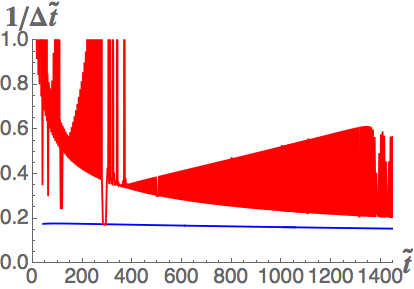}
\caption{\small \baselineskip 11pt The change in the effective local frequencies, $1 / \Delta \tilde{t}_i$, as a function of $\tilde{t} = \sqrt{\lambda_\phi} \, M_{\rm pl} \, t / \xi_\phi$, for the background inflaton field $\phi (t)$ (blue) and for the isocurvature mode $z_k (t)$ (with $k = 0$, red), for different values of $\xi_\phi$: $\xi_\phi = 10^2$ ({\it left}), and $\xi_\phi = 10^3$ ({\it right}). }
\label{fig:broadnarrow}
\end{figure}

We confirm the behavior, and the scaling with $\tilde{t}$, in Fig.~\ref{fig:broadnarrow}. For these plots, we calculate all zero-crossings of $\phi (t)$ and of $z_k (t)$ (for $k = 0$). For each field, we denote $\Delta \tilde{t}_i$ as the difference in time between the $i$th and $(i + 1)$th zero-crossings. The inverse of this quantity, $1 / \Delta \tilde{t}_i$, may be considered a local measure of frequency, which acquires different values for $\phi (t)$ and $z_k (t)$. Fig.~\ref{fig:broadnarrow} shows the evolution of these local frequencies for the background field and the isocurvature mode. As expected, the frequency of $\phi (t)$ does not show significant variation, whereas the local frequency of $z_k (t)$ displays significant dispersion, due to the strongly anharmonic oscillation of $m_{\rm eff,\chi}^2$. Because of the structure of $m_{\rm eff,\chi}^2$, the local frequency for $z_k (t)$ is comprised of both short- and long-duration components. Whereas at early times, the effective local frequency for $z_k (t)$ is considerably greater than for $\phi (t)$, the two become comparable around $\tilde{t} \sim 130$ for $\xi_\phi = 10^2$, and around $\tilde{t} \sim 1300$ for $\xi_\phi = 10^3$, in keeping with our expectations from the scaling arguments given above.

\begin{figure}
\includegraphics[width=0.5\textwidth]{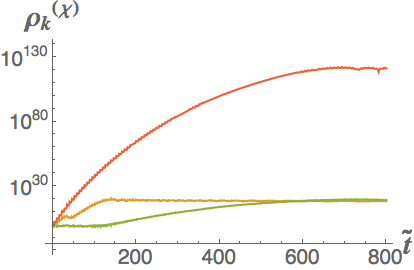}
\caption{\small \baselineskip 11pt The energy density of isocurvature modes $\rho_k^{(\chi)}$ versus $\tilde{t} = \sqrt{\lambda_\phi} \, M_{\rm pl} \, t / \xi_\phi$ for $k = 0$ and $\xi_\phi = 100$, calculated based on $m_{2,\chi}^2$ alone (orange), on $m_{1,\chi}^2$ alone (green), and on $m_{\rm eff,\chi}^2$ (yellow), for the same ratios of couplings as in Fig.~\ref{fig:allresultsintro}. }
\label{fig:isonog}
\end{figure}

Another way to identify the importance of $m_{2,\chi}^2$ in driving the early amplification of isocurvature modes is to calculate $\rho_k^{(\chi)}$ by neglecting either $m_{1,\chi}^2$ or $m_{2,\chi}^2$. The results are shown in Fig.~\ref{fig:isonog} for $\xi_\phi = 100$. If there were no contributions from the potential --- that is, if we set $m_{1,\chi}^2 = 0$, which is akin to setting $g = 0$ so that there were no direct coupling between $\phi$ and $\chi$ --- one would still find the same initial burst of growth in $\rho_k^{(\chi)}$, stemming entirely from the curved field-space manifold (which itself arises from the nonminimal couplings upon transforming to the Einstein frame). On the other hand, if one neglected the effects of $m_{2,\chi}^2$ and only considered the role of $m_{1,\chi}^2$, then one would find little initial growth. The nonminimal couplings thus open a distinct ``channel" for $\phi$ and $\chi$ to interact, via their shared coupling to the spacetime Ricci scalar, $R$, in the Jordan frame \cite{BassettGeometric,ShinjiPapers}. In the limit of large couplings, $\xi_I \geq {\cal O} (10^2)$, this channel can drive broad resonances.

\begin{figure}
\centering
\includegraphics[width=.48\textwidth]{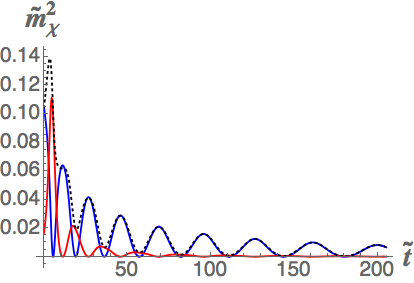} 
  \caption{\small \baselineskip 11pt
The rescaled effective mass of the isocurvature perturbations $\tilde{m}_{\rm eff,\chi}^2 = \xi_\phi m_{\rm eff,\chi}^2$ (black-dotted) as a function of $\tilde{t} = \sqrt{\lambda_\phi} \, M_{\rm pl} \, t / \xi_\phi$ for $\xi_\phi = 1$. Also shown are $\tilde{m}_{1,\chi}^2$ (blue) and $\tilde{m}_{2,\chi}^2$ (red). }
\label{fig:mchi_k1}
\end{figure}

We conclude this section by considering the case of $\xi_\phi \leq 1$, which departs strongly from both the intermediate- and large-$\xi_I$ regimes. From Fig.\ \ref{fig:mchi_k1} we note that the contribution to $m_{\rm eff,\chi}^2$ arising from the curved field-space manifold, $m_{2,\chi}^2$, becomes subdomiant after the first oscillation of the background field. In keeping with our analysis of the small-$\xi_I$ regime in the rigid-spacetime approximation \cite{MultiPreheat2}, we find that the effects of the curved field-space manifold remain modest for $\xi_I \leq {\cal O} (1)$, making little change from the minimally coupled case.

\section{Results: Nonzero wavenumber}
\label{sec:Nonzerok}

In Sections \ref{sec:Adiabatic} and \ref{sec:Isocurvature} we analyzed the behavior of super-horizon modes with $k / aH \ll 1$, focusing on the limit $k \rightarrow 0$. In this section we briefly examine the behavior of adiabatic and isocurvature modes with nonzero wavenumber, for modes with wavenumber comparable to the Hubble radius ($k / aH \sim 1$) as well as for modes whose wavenumber begins well within the Hubble radius ($k / aH \gg 1$) at the start of preheating. We restrict attention to the large-$\xi_I$ regime, in which both adiabatic and isocurvature modes with $k = 0$ are amplified efficiently.

\begin{figure}
\centering
\includegraphics[width=.48\textwidth]{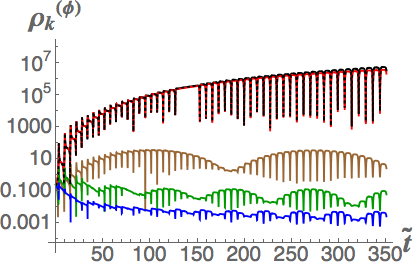} \quad \includegraphics[width=.48\textwidth]{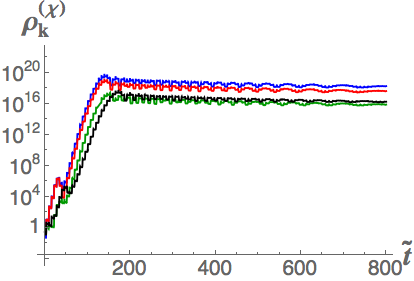} 
  \caption{\small \baselineskip 11pt ({\it Left}) The energy density for adiabatic modes versus $\tilde{t} = \sqrt{\lambda_\phi} \, M_{\rm pl} \, t / \xi_\phi$ with $\xi_\phi=100$ and $k/ [a(t_{\rm end})H(t_{\rm end})] = 0,1,5,10,20$ (black-dashed, red, brown, green, and blue, respectively). ({\it Right}) The energy density for isocurvature modes with $\xi_\phi = 100$ and $k / [a (t_{\rm end}) H (t_{\rm end})] = 0, 2, 10, 20$ (blue, red, green, and black, respectively). For both plots we use $\xi_\chi / \xi_\phi = 0.8$, $\lambda_\chi / \lambda_\phi = 1.25$, and $g / \lambda_\phi = 1$. }
\label{fig:kne0}
\end{figure}

Fig.~\ref{fig:kne0} shows the behavior of $\rho_k^{(\phi)}$ and $\rho_k^{(\chi)}$ for values of $k$ within the range $0 \leq k / [a(t_{\rm end} ) H (t_{\rm end}) ] \leq 20$. For the adiabatic perturbations, we find a sharp distinction between sub-horizon and super-horizon modes: only modes with $k / a H \leq 1$ at the start of preheating become amplified. This behavior is in accord with our findings in Ref.~\cite{MultiPreheat2}: for $\xi_I \geq {\cal O} (10^2)$, the dominant resonance band for adiabatic modes occurs for $k \rightarrow 0$, while resonance bands for larger $k$ shrink to the narrow-resonance regime (and hence do not remain very effective in an expanding universe). Moreover, as we found in Section \ref{sec:Adiabatic}, in a dynamical spacetime the dominant amplification for adiabatic modes at early times stems from effects of the coupled metric perturbations, which drive tachyonic growth with $m_{\rm eff,\phi}^2 < 0$. Such tachyonic growth will only be effective for modes with $(k / a)^2 < m_{\rm eff,\phi}^2$, such that the effective frequency becomes imaginary, $\Omega_{(\phi)}^2 (k,t) / a^2 = (k / a)^2 + m_{\rm eff,\phi}^2 < 0$. Sub-horizon modes, with $k / aH \gg 1$ at the start of inflation, therefore do not undergo the initial tachyonic amplification. Meanwhile, the early burst of tachyonic amplification (for super-horizon modes) persists only for as long as $m_{3,\phi}^2$ dominates $m_{\rm eff,\phi}^2$, which scales (as we found in Section \ref{sec:Adiabatic}) as $\tilde{t} \sim \sqrt{\xi_\phi}$. By the time the wavelengths of sub-horizon modes redshift to super-horizon scales, the tachyonic amplification typically has ceased.

The behavior of the isocurvature modes is quite different. In Ref.~\cite{MultiPreheat2} we found that for $\xi_I \geq {\cal O} (10^2)$, the Floquet charts for the isocurvature modes show dense bands of instability regions, with little $k$-dependence across a wide range of $k$. Moreover, as discussed in Section \ref{sec:Isocurvature}, the growth of long-wavelength isocurvature perturbations occurs in the broad-resonance regime at early times, when the modes $z_k (t)$ oscillate more rapidly than the inflaton $\phi (t)$. The effective frequency for modes $z_k (t)$ increases with larger $k$, and the broad-resonance instabilities identified in Ref.~\cite{MultiPreheat2} remain robust in an expanding spacetime. As shown in Fig.~\ref{fig:kne0}, the spectrum of $\rho_k^{(\chi)}$ therefore remains nearly flat for super-horizon and sub-horizon modes.

\section{Observational Consequences}
\label{sec:Observational}

In general for multifield models, the growth of long-wavelength isocurvature perturbations can affect the evolution of super-horizon adiabatic modes \cite{WandsReview,MazumdarRocher,MartinRingeval,VenninWands,GongMultifield,GWBM,Groot,Elliston:2011dr, Leung:2012ve,ChambersRajantie,BondFrolov,Bethke}. Such transfers of power from isocurvature to adiabatic perturbations can affect CMB observables, such as the spectral index $n_s$ or primordial non-Gaussianity $f_{\rm NL}$. In some cases, this process can produce significant corrections from the preheating phase to values of observables calculated during inflation. (See the discussion in Section 7 of Ref.~\cite{AHKK}, and references therein.) 

Although we have found very efficient growth of isocurvature modes on super-horizon scales during preheating for $\xi_I \geq {\cal O} (10^2)$, such growth need not pose a direct threat to the close fit between the predictions for primordial observables such as $n_s$, $r$, and $f_{\rm NL}$ from these models and the latest observations \cite{KMS,GKS,KS,SSK}. In Ref.~\cite{KMS} we examined the evolution of the gauge-invariant curvature and isocurvature perturbations, ${\cal R}_c$ and ${\cal S}$, finding, in the long-wavelength limit:
\beqn
\begin{split}
\dot {\cal R}_c &= 2 \omega {\cal S} + {\cal O}\left ({k^2 \over a^2 H^2} \right ) \\
\dot {\cal S} &= \beta H {\cal S} + {\cal O}\left ({k^2 \over a^2 H^2} \right ) ,
\end{split}
\label{RdotSdot}
\eeqn
where $\omega(t)$ is the turn-rate of the fields' trajectory in field-space, and $\beta(t)$ encodes the effects of $m_{\rm eff,\chi}^2$. For motion within a single-field attractor the turn-rate vanishes identically, $\omega = 0$, and hence the gauge-invariant curvature perturbation ${\cal R}_c$ remains conserved in the long-wavelength limit. We found in Ref.~\cite{MultiPreheat1} that the single-field attractor behavior persists in these models throughout preheating as well as during inflation. Thus, at least at the level of a linearized analysis, we do not anticipate significant effects on primordial observables from the amplification of $\rho_k^{(\chi)}$ after the end of inflation. 

Nonlinear effects in such models could certainly become important after the end of inflation, such as the formation of oscillons, which could affect the evolution of the effective equation of state and thereby the expansion history \cite{GleiserOscillons,AminOscillons,ZhouOscillons,Mustafa,MustafaDuration}. Moreover, the rapid growth of $\rho_k^{(\chi)}$ in these models suggests that the majority of the inflaton's energy could be transferred to isocurvature modes within a single background oscillation \cite{Ema:2016dny}. Recent lattice simulations of related models that yield near-instantaneous preheating have identified qualitative differences from linear analyses, such as significant re-scattering between the inflaton and produced particles \cite{Adshead:2015pva,Adshead:2016iae,Child:2013ria}. Such inherently nonlinear effects are beyond the scope of the present analysis.

\section{Potential Implications for Higgs Inflation}
\label{sec:Higgs}
 
The strong resonances that we have identified in multifield models with $\xi_I \gg 1$ have potential implications for reheating after Higgs inflation \cite{BezrukovShaposhnikov}. Most studies of Higgs inflation have adopted the unitary gauge, in which the Higgs sector reduces to a single scalar degree of freedom. However, it is well known that the unitary gauge suffers from poor behavior at energies that are large compared to the symmetry-breaking scale \cite{WeinbergQFT}. The so-called $R_\xi$ gauges, on the other hand, remain well behaved even for energies well above the symmetry-breaking scale. In the $R_\xi$ gauges (which include the Feynman - 't Hooft gauge and the Landau gauge), the Goldstone modes remain in the spectrum \cite{WeinbergQFT,Barvinsky,Burgess,HertzbergNonmin,Mooij}. In the framework we have developed here, these cases correspond to symmetric couplings among the scalar fields ($\lambda_\phi = \lambda_\chi = g$ and $\xi_\phi = \xi_\chi$), and the Goldstone modes correspond to isocurvature perturbations \cite{GKS,KaiserTodhunter}.

A salient feature of the $R_\xi$ gauges is the Goldstone boson equivalence theorem \cite{equivtheorem}, according to which the high-energy dynamics of electroweak vector gauge bosons, such as the amplitudes for $W$ and $Z$ scattering, is captured by the dynamics of the Goldstone and Higgs scalars. This important result follows from the fact that the longitudinal-polarization vectors for the gauge bosons scale with momentum whereas the transverse-polarization states do not. Thus the amplitudes for high-energy scattering processes are dominated by the longitudinal polarizations of the vector bosons. And the longitudinal polarizations, in turn, are related to the Goldstone bosons via spontaneous symmetry breaking. Thus one may calculate such quantities as the amplitude for $W^\mu W^\mu$ scattering or $W^\mu$-Higgs scattering in terms of the simpler interactions among the Goldstone and Higgs scalars \cite{equivtheorem}.

The Goldstone boson equivalence theorem stipulates that amplitudes for the high-energy scattering of vector gauge bosons are {\it identical} to corresponding amplitudes involving the Goldstone and Higgs scalars, up to corrections of order ${\cal O} (m_W / \sqrt{s})$, where $m_W$ is the mass of the $W$ boson and $\sqrt{s}$ is the center-of-mass energy for a given process \cite{equivtheorem}. During and after inflation, the Higgs field has a (time-dependent) vacuum expectation value, $h (t) \sim \langle \phi (t) \rangle,$ since the background field $\phi$ is displaced from the minimum of its potential. Therefore the electroweak symmetry is broken during and after inflation, and should remain broken until the Higgs condensate dissolves at the end of reheating. (See also Refs.~\cite{HiggsReheat1,HiggsReheat2,HiggsReheat3,HiggsReheat4}.) In the broken-symmetric phase, $m_W =  \tilde{g} \, h / 2$, where $\tilde{g} \sim {\cal O} (1)$ is the electroweak gauge coupling. During reheating, we then have $m_W \sim h$, and the equivalence theorem holds for energies $\sqrt{s} > h$. 

At the start of preheating, the value of the vacuum expectation value is given by the dynamics of the background field $\phi (t)$:
\beq
h \sim \langle \phi \rangle \lesssim \frac{ M_{\rm pl} }{\sqrt{\xi_\phi } } .
\label{hvev}
\eeq
During reheating, the amplitude of the Higgs condensate decays due to the expansion of the universe (even in the linearized approximation), and also due to the transfer of energy to fluctuations (which requires a nonlinear analysis). For the scalar sector, meanwhile, the strong-coupling scale in the Einstein frame, $\Lambda_E$, as calculated in Refs.~\cite{BezrukovConsistency,BezrukovDilaton,BezrukovInflaton}, satisfies $(h / \Lambda_E) < 1$ for all values $0 < h \leq M_{\rm pl } / \sqrt{\xi_\phi}$. Therefore the Goldstone boson equivalence theorem should be relevant during preheating, for high-energy processes with $h < \sqrt{s} < \Lambda_E$.

To match the COBE normalization for the amplitude of primordial perturbations in this model, we require \cite{HiggsReheat1,BezrukovTwoLoop}
\beq
\frac{\xi_\phi}{\sqrt{\lambda_\phi} } = 4.7 \times 10^4 \, .
\label{xilambdaCOBE}
\eeq
From Eq.~(\ref{Hend}) we then find
\beq
H_{\rm end} \simeq  0.4 \sqrt{ \frac{ \lambda_\phi  }{12 \xi_\phi^2 } } \, M_{\rm pl} = 2.5 \times 10^{-6} \, M_{\rm pl} = 6.0 \times 10^{12} \, {\rm GeV}.
\label{HendHiggs}
\eeq
For best-fit values of the masses of the Higgs boson and the top quark, the running of the Higgs self-coupling $\lambda_\phi$ yields $\lambda_\phi (\mu) \simeq 5.0 \times 10^{-3}$ at the energy scale $\mu = H_{\rm end} = 6 \times 10^{12}$ GeV \cite{BezrukovTwoLoop}. Eq.~(\ref{xilambdaCOBE}) then suggests that $\xi_\phi \simeq 3.3 \times 10^3$ at the start of preheating. (If the top-quark mass is greater than the current best-fit value, then one expects $\lambda_\phi (\mu) < 5.0 \times 10^{-3}$ and hence $\xi_\phi < 3.3 \times 10^3$ at $\mu = 6 \times 10^{12}$ GeV \cite{Allison13,BezrukovTopQuark}.) Given the efficient amplification of isocurvature modes in such models with $\xi_\phi \sim 10^3$, we may expect preheating after Higgs inflation to conclude within a few oscillations of the inflaton. 

The nonzero value of $h = \langle \phi \rangle$ induces a mass for the Goldstone bosons $\chi$. We may estimate the mass-scale due to electroweak symmetry-breaking at the start of preheating as
\beq
m_{\rm vev}^2 \sim \lambda_\phi h^2 \sim \frac{ \lambda_\phi M_{\rm pl}^2}{\xi_\phi } .
\label{mvev}
\eeq
Given Eq.~(\ref{HendHiggs}), we then find
\beq
\frac{ m_{\rm vev}^2 }{H_{\rm end}^2 } \sim 100\, \xi_\phi \, .
\label{mvevHend}
\eeq
Meanwhile, in Eq.~(\ref{m2Hcompare}) we calculated the ratio of the Hubble scale to the spike in $m_{\rm eff,\chi}$ due to the nontrivial field-space manifold, $m_{2,\chi}$, and found
\beq
\frac{ m_{2, \chi}^2 }{H^2} \sim \xi_\phi^2 \, , 
\label{m2Hb}
\eeq
so for $\xi_\phi \sim 10^3$, we expect $m_{2,\chi} \gg m_{\rm vev}$ at the start of preheating, since
\beq
\frac{ m_{2, \chi}^2 }{m_{\rm vev}^2 } \sim \frac{\xi_\phi }{100} .
\label{m2mvev}
\eeq
Even in the presence of electroweak symmetry breaking during the preheating epoch, the features that are most responsible for rapid particle production --- namely, the spikes in $m_{2,\chi}^2$ which drive violations of the adiabatic condition --- dominate the dynamics of the scalar sector.

Significant backreaction from created particles will halt the background field's oscillations, with $\phi$ settling near $\phi \sim 0$, leaving a universe filled with scalar Higgs bosons. (See also Refs.~\cite{Ema:2016dny,HiggsReheat1,HiggsReheat2,HiggsReheat3,HiggsReheat4}.) We found in Ref.~\cite{MultiPreheat1} that the effective equation of state after inflation tends toward $w_{\rm avg} \sim 1/3$ within several efolds after the end of inflation for $\xi_I \geq 100$, based on analysis of the background field's dynamics. Since the Higgs potential is dominated by the quartic term after the end of inflation, the Higgs bosons produced during preheating should behave as nearly massless radiation modes (especially once $\phi \sim 0$), quickening the rate at which $w_{\rm avg} \rightarrow 1/3$. Thus we expect a rapid transition to a radiation-dominated equation of state after the end of inflation.

The successes of big-bang nucleosynthesis require more than just a radiation-dominated equation of state; the universe must also become filled with a plasma of Standard Model particles some time after inflation ends \cite{AHKK}. Ref.~\cite{Adshead:2016iae} studied the generation of a charged plasma from instantaneous preheating. In that scenario, the universe becomes filled with hypercharge gauge bosons after inflation, which scatter into Standard Model particles. In the case of Higgs inflation, we may expect an inverse process to unfold.

\begin{figure}
\centering
\includegraphics[width=.3\textwidth]{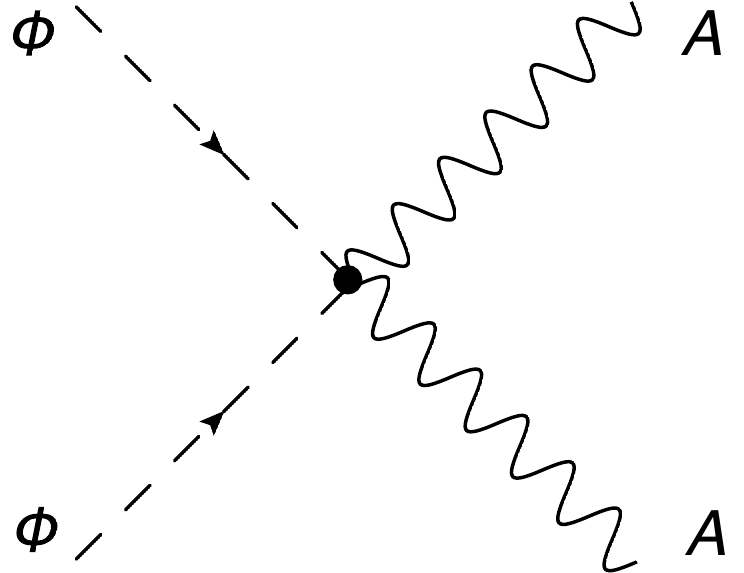} \hspace{0.1\textwidth} \includegraphics[width=.3\textwidth]{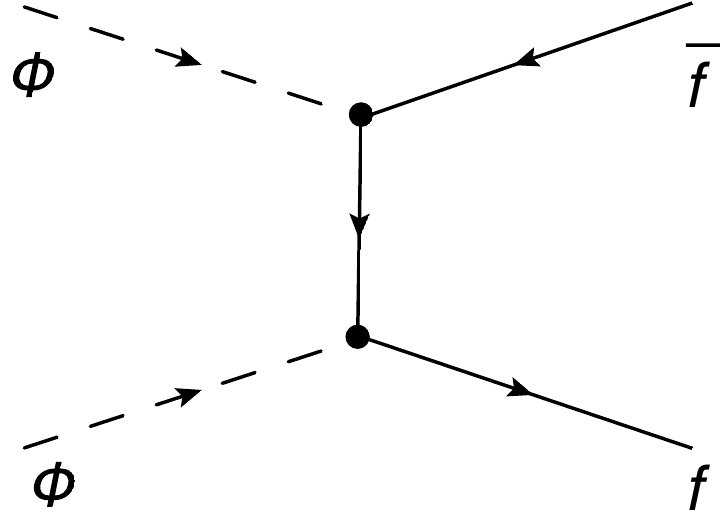} 
  \caption{ \small \baselineskip 11pt
Scattering vertices that allow the transfer of energy from the Higgs bosons to the rest of the Standard Model, specifically into gauge fields ({\it left}) and fermions ({\it right}). }\label{fig:higgsdecay}
\end{figure}

The energy density of the universe at the end of inflation is $\rho \sim M_{\rm pl}^2 H_{\rm end}^2$, while the momentum of each Higgs mode can be taken to be comparable to the Hubble scale at the end of inflaton, $E\sim k/a_{\rm end} \sim H_{\rm end}$. Thus the particle density scales as $n = \rho / E \sim M_{\rm pl}^2 H_{\rm end}$, while the velocity for radiation modes is simply $v\sim c$. The speed of the transition from a Higgs boson bath into a plasma depends on the ratio of the scattering rate to the Hubble expansion rate,
\beq
{\Gamma \over H } = {n \sigma v \over H} .
\label{eq:gammaoverH}
\eeq
The scattering rate of Higgs particles into gauge bosons or fermions proceeds through vertices like the ones shown in Fig. \ref{fig:higgsdecay}, and scales as 
\beq
\sigma \sim {\alpha_Y^2 \over s} \sim  {\alpha_Y^2 \over H_{\rm end}^2} 
\eeq
for the case of scattering into hypercharge $U(1)$ bosons. Scattering into fermions has a similar cross-section, and thus we may estimate for Eq.~(\ref{eq:gammaoverH}), 
\beq
{\Gamma \over H }  \sim     a_Y^2 \left ( {M_{\rm pl} \over  H_{\rm end}} \right )^2 \sim 10^9 a_Y^2 \gg 1 .
\eeq
We therefore expect that scattering of the Higgs bosons into the rest of the Standard Model particles should occur almost instantaneously. Thermalization would proceed through similar interactions, and should also be efficient. After the electroweak phase transition (which presumably would occur at much lower energies than either inflation or reheating), the Higgs field would acquire a nonzero vacuum expectation value of $v\approx 246\,\rm{ GeV}$. Then any remaining massive Higgs particles would decay into Standard Model particles, following the experimentally known decay rates \cite{Olive:2016xmw,SMfn}.

In sum, for the case of reheating after Higgs inflation, with $\xi_\phi \gtrsim10^3$, we may expect that the efficient production of isocurvature modes --- which, in this case, correspond to Goldstone modes within the Standard Model Higgs doublet and therefore behave as the longitudinal modes of the $W$ and $Z$ bosons when electroweak symmetry is broken --- would lead to the rapid break-up of the Higgs condensate into radiation modes, followed by a fast transition to a primordial plasma of Standard Model particles. In such a scenario, the reheat temperature after inflation could be as high as $T_{\rm reh} \sim \sqrt{H_{\rm end} M_{\rm pl}} \sim 10^{-3} M_{\rm pl}$.

Such a high reheat temperature may pose a challenge for a more thorough investigation of reheating after Higgs inflation, beyond the linearized analysis of the preheating phase we have pursued here. Typical momenta for quanta in a blackbody distribution scale as $k \sim 3T$. So if the universe achieves thermal equilibrium at a temperature as high as $T_{\rm reh} \sim 10^{-3}\, M_{\rm pl}$ after inflation, then typical scatterings among particles would involve momentum exchanges of order $k \sim 3 T_{\rm reh}$. Meanwhile, in the scenario we have described here, the inflaton condensate would likely dissipate quickly, falling from $h \sim M_{\rm pl} / \sqrt{\xi_\phi}$ to $h < M_{\rm pl} / \xi_\phi$ via various nonlinear processes within a few oscillations. Following Refs.~\cite{BezrukovConsistency,BezrukovDilaton,BezrukovInflaton}, one may estimate the perturbative unitarity cut-off scale in the Einstein frame, $\Lambda_E$, arising from tree-level interactions within the scalar sector. The cut-off scale depends on $h = \langle \phi \rangle$; one finds that $\Lambda_E$ falls to $\Lambda_E \sim M_{\rm pl} / \xi_\phi$ for $h < M_{\rm pl} / \xi_\phi$. In that case, typical momentum exchanges from particle scatterings within the thermal plasma would be of order
\beq
\frac{k}{\Lambda_E} \sim \frac{ 3 T_{\rm reh} }{M_{\rm pl} /\xi_\phi} \sim 3 \, \xi_\phi \times 10^{-3} \sim {\cal O} (10) 
\label{LambdaTreh}
\eeq
for $\xi_\phi \sim 3 \times 10^3$. Such a large ratio of $k / \Lambda_E$ strongly suggests that some portion of the reheating dynamics in such a scenario would fall within the strong-coupling regime; studying such dynamics may well prove intractable. (The ratio in Eq.~(\ref{LambdaTreh}) remains unchanged if one works in unitary gauge and calculates the strong-coupling scale in the Einstein frame from perturbative scattering of the Higgs field and vector gauge bosons, $\Lambda_{\rm gauge}$, since, for $h < M_{\rm pl} / \xi_\phi$, one has $\Lambda_E \sim \Lambda_{\rm gauge} \sim M_{\rm pl} / \xi_\phi$ \cite{BezrukovConsistency,BezrukovDilaton,BezrukovInflaton}.)  Meanwhile, given $h < \Lambda_E$, we similarly estimate $k / h \geq {\cal O} (10)$ in such a scenario, further underscoring the appropriateness of renormalizable $R_\xi$ gauges for analyzing the post-inflation dynamics.

Addressing the late-stage thermalization following the initial phase of preheating remains beyond the scope of this paper, and deserves further study. One interesting question relevant to such further study concerns the behavior of the unitarity cut-off scale itself. The estimate $\Lambda_E \sim M_{\rm pl} / \xi_\phi$ comes from analyzing perturbative processes involving a few particles in both the incoming and outgoing states, such as $2 \rightarrow 2$ scattering \cite{BezrukovConsistency,BezrukovDilaton,BezrukovInflaton}. But preheating involves the rapid production of {\it many} particles in the final state, with number densities $n_k \sim \exp [ 2 \mu_k t ] \gg 1$. Even for perturbative analyses, if the final state involves $n \gg 1$ bosons, then phase-space factors generically raise $\Lambda_{\rm pert}^{(n)}$ above what one would calculate for $2 \rightarrow 2$ scattering, with $\Lambda_{\rm pert}^{(n)}$ typically growing linearly with $n$ in the limit $n \gg 1$ \cite{DicusHeBelyaev}. Moreover, the production of particles during preheating involves inherently {\it nonperturbative} processes. The cross section for nonperturbative processes that yield $n$-particle final states, $\sigma_{1 \rightarrow n}$, has a very different scaling with energy than the corresponding perturbative cross sections for {\it few} $\rightarrow$ {\it few} scattering \cite{VoloshinNonPert}. If these considerations were to yield a cut-off scale considerably different than $\Lambda_E$, then ratios like Eq.~(\ref{LambdaTreh}) would need to be revisited. Such questions remain the topic of further research.

\section{Conclusions}
\label{sec:Conclusions}

Building on Refs.~\cite{MultiPreheat1,MultiPreheat2}, we have studied the preheating dynamics of adiabatic and isocurvature perturbations in multifield models with nonminimal couplings. We exploit a doubly-covariant framework and work to linear order in the perturbations, including both the expansion of the universe during preheating and the effects of the coupled metric perturbations. We focus on the dynamics after inflation when the fields evolve within a single-field attractor, which is a generic feature of this class of models.

We find that the behavior of the adiabatic and isocurvature perturbations depends on distinct sets of contributions to their effective masses, $m_{ {\rm eff}, I}^2$. The effective mass for the adiabatic modes, $m_{\rm eff,\phi}^2$, is dominated by a term arising from the potential, $m_{1,\phi}^2$, and a term arising from the coupled metric perturbations, $m_{3,\phi}^2$. At early times, the metric perturbations dominate, driving a broad, tachyonic amplification of super-horizon modes for $\xi_I \geq {\cal O} (100)$. The time-scale during which the tachyonic amplification persists scales as $\tilde{t} \sim \sqrt{\xi_\phi}$, where $\tilde{t} = \sqrt{\lambda_\phi} \, M_{\rm pl} \,  t / \xi_\phi \propto H (t_{\rm end} ) t$. Sub-horizon modes of the adiabatic perturbations are not affected by the tachyonic instability, and in general are not strongly amplified after inflation.

The behavior of the isocurvature perturbations is driven by a different pair of contributions to the effective mass, $m_{\rm eff,\chi}^2$, namely, $m_{1,\chi}^2$ from gradients of the potential and $m_{2,\chi}^2$ from the curved field-space manifold (which itself arises from the nonminimal couplings, upon transforming to the Einstein frame). In contrast to the adiabatic modes, the amplification of isocurvature modes is efficient for sub-horizon and super-horizon scales alike, and their growth rate for $\xi_I \geq {\cal O} (100)$ exceeds that of the super-horizon adiabatic modes. The duration of efficient, broad-resonance amplification for the isocurvature modes scales as $\tilde{t} \sim \xi_\phi$.

The amplification of both adiabatic and isocurvature modes becomes more efficient as the nonminimal couplings become large, $\xi_I \geq {\cal O} (100)$, underscoring the role of the nonminimal couplings in driving the dynamics. The adiabatic modes grow primarily due to their coupling to metric perturbations; that coupling increases as $\xi_I$ increases, which is most clear in the Jordan frame, given the coupling $f (\phi^I) R$ in the action. Likewise, the isocurvature modes grow primarily due to the nontrivial field-space manifold, which is likewise a manifestation of the underlying nonminimal couplings, and whose curvature increases with $\xi_I$. The curvature of the field space manifold, ${\cal R}_{ILMJ}$, allows for efficient transfer of energy from the inflaton condensate to isocurvature perturbations, even in the absence of a direct coupling (within the potential) between the various fields. 

Within the scope of our linearized analysis, we do not expect the rapid amplification of super-horizon isocurvature modes to affect inflationary predictions for primordial observables (such as the spectral index or non-Gaussianity), because the single-field attractor prohibits isocurvature modes from sourcing a change in the gauge-invariant curvature perturbation on long wavelengths. Whether such behavior persists beyond linear order in the perturbations remains the subject of further study. In the meantime, the rapid amplification of isocurvature modes suggests that reheating in models like Higgs inflation, with $\xi_\phi \sim 10^3$, could be especially efficient.

\section*{Appendix: Comparing Relevant Time-Scales}

In Section \ref{sec:Adiabatic}, we found that the effective mass for the adiabatic modes, $m_{\rm eff,\phi}^2$, is dominated by two contributions: $m_{1,\phi}^2$, arising from gradients of the Einstein-frame potential, and $m_{3,\phi}^2$, arising from the coupled metric perturbations. In particular, the effects of $m_{3,\phi}^2$ dominate the dynamics at early times, while $m_{1,\phi}^2$ dominates at later times. In terms of the rescaled time coordinate, 
\beq
 \tilde{t} \equiv \sqrt{ \frac{\lambda_\phi}{\xi_\phi^2} } \, M_{\rm pl} \, t  \, ,
\label{tildetdef}
\eeq
we may label the cross-over time between these two regimes as $\tilde{t}_{\rm cross}^{(\phi)}$. In Section \ref{sec:Adiabatic} we found that $\tilde{t}_{\rm cross}^{(\phi)}$ scales as
\beq
\tilde{t}_{\rm cross}^{(\phi)} \sim \sqrt{\xi_\phi}
\label{tcrossphi}
\eeq
in the limit $\xi_\phi \gg 1$, for evolution within a single-field attractor along the direction $\chi \sim 0$.

In Section \ref{sec:Isocurvature}, meanwhile, we found that the effective mass for the isocurvature modes, $m_{\rm eff,\chi}^2$, is dominated by a different pair of contributions: $m_{1,\chi}^2$, arising from gradients of the Einstein-frame potential, and $m_{2,\chi}^2$, arising from the nontrivial field-space manifold. In the case of the isocurvature perturbations, the effects of $m_{2,\chi}^2$ dominate at early times. In terms of $\tilde{t}$, we found the cross-over time between the two regimes for the isocurvature modes to scale as 
\beq
\tilde{t}_{\rm cross}^{(\chi)} \sim \xi_\phi
\label{tcrosschi}
\eeq
in the limit $\xi_\phi \gg 1$, again for evolution within a single-field attractor along the direction $\chi \sim 0$.

We may compare the scaling of $\tilde{t}_{\rm cross}^{(\phi)}$ and $\tilde{t}_{\rm cross}^{(\chi)}$ with that of a third time-scale, namely, the cross-over time in the equation of state for the background dynamics, $w_{\rm avg}$, between a matter-dominated and radiation-dominated phase. We found in Ref.~\cite{MultiPreheat1} that $w_{\rm avg} \sim 0$ at the start of preheating in this family of models, and that $w_{\rm avg} \rightarrow 1/3$ within several efolds after the end of inflation, though the number of efolds until $w_{\rm avg} \sim 1/3$ grew with $\xi_\phi$. (See Fig.~9 in Ref.~\cite{MultiPreheat1}.) Here we estimate $\tilde{t}_{\rm cross}^{(w)}$, the cross-over time between matter-dominated and radiation-dominated evolution of the background dynamics.

Within a single-field attractor along $\chi \sim 0$, the Einstein-frame potential simplifies to
\beq
V(\phi,\chi) \simeq \frac{ \lambda_\phi \, M_{\rm pl}^4 }{4} \frac{\phi^4}{[ M_{\rm pl}^2 + \xi_\phi \phi^2 ]^2 } = \frac{\lambda_\phi \, M_{\rm pl}^4}{4 \xi_\phi^2} \frac{ \delta^4}{ [1 + \delta^2 ] ^2 } ,
\label{Vsfa}
\eeq
where, as usual, we define $\delta \equiv \sqrt{ \xi_\phi } \, \phi / M_{\rm pl}$. Inflation ends at $\delta \lesssim 1$, and the amplitude of $\phi$ continues to fall as preheating proceeds. For $\delta < 1$, we may therefore approximate
\beq
V \simeq \frac{\lambda_\phi \, M_{\rm pl}^4 }{4 \xi_\phi^2} \delta^4 .
\label{Vapprox}
\eeq
At early times during preheating, the time-averaged equation of state obeys $w_{\rm avg} \simeq 0$, and hence the energy density scales as
\beq
\rho \simeq \rho_0 e^{-3 N} ,
\label{rhoN}
\eeq
where $N$ is the number of efolds since the end of inflation. We may estimate
\beq
\rho \simeq V_{\rm max} = \frac{\lambda_\phi \, M_{\rm pl}^4 }{4 \xi_\phi^2} \delta_{\rm max}^4 ,
\label{rhomax}
\eeq
where $\delta_{\rm max} = \sqrt{\xi_\phi} \, \phi_{\rm max} / M_{\rm pl}$ is the rescaled amplitude of the field's oscillations. If we label $\alpha \equiv \sqrt{\xi_\phi} \, \phi_0 / M_{\rm pl}$, where $\phi_0$ is the initial amplitude of $\phi (t)$ at the start of preheating, then from Eqs.~(\ref{rhoN}) and (\ref{rhomax}) we may estimate
\beq
\delta_{\rm max} (t) \simeq \alpha e^{-3 N / 4} .
\label{deltamaxN}
\eeq
Next we may estimate
\beq
H^2 \simeq  \frac{1}{ 3 M_{\rm pl}^2} V_{\rm max} = \frac{ \lambda_\phi \, M_{\rm pl}^2}{ 12 \xi_\phi^2} \delta_{\rm max}^4 (t) ,
\label{Hdeltamax}
\eeq
with which we may write
\beq
dN = H dt \simeq \frac{ M_{\rm pl} }{2 \sqrt{3} } \sqrt{ \frac{\lambda_\phi} {\xi_\phi^2} } \, \delta_{\rm max}^2 (t) dt ,
\label{dN}
\eeq
or, upon using Eqs.~(\ref{tildetdef}) and (\ref{deltamaxN}),
\beq
 e^{3N / 2} \simeq \frac{ \sqrt{3}}{4} \alpha^2  \, \tilde{t} .
\label{Ntildet1}
\eeq

As we saw in Section IIIC of Ref.~\cite{MultiPreheat1}, the evolution of the background-order equation of state depends on the shifting balance between the kinetic energy, $\dot{\sigma}^2$, and the potential, $V (\phi^I)$. Within the single-field attractor, we have
\beq
\dot{\sigma}^2 \equiv {\cal G}_{IJ} \dot{\varphi}^I \dot{\varphi}^J \rightarrow {\cal G}_{\phi\phi} \dot{\phi}^2 ,
\label{dotsigma1}
\eeq
where
\beq
{\cal G}_{\phi\phi} = \left( \frac{ M_{\rm pl}^2}{2 f} \right) \left[ 1 + \frac{ 3 \xi_\phi^2 \phi^2}{f} \right] = \left( \frac{1}{1+ \delta^2} \right) \left[ \frac{ 1 + \delta^2 + 6 \xi_\phi \delta^2 }{1 + \delta^2 } \right] .
\label{Gphiphi1}
\eeq
At the start of preheating, we have $\delta \sim {\cal O} (1)$, and hence ${\cal G}_{\phi\phi} \sim \xi_\phi \delta^2 \gg 1$ for $\xi_\phi \gg 1$. However, ${\cal G}_{\phi\phi}$ falls during preheating as the amplitude of $\phi (t)$ falls, eventually reaching ${\cal G}_{\phi\phi} \rightarrow 1$ for $6 \xi_\phi \delta^2 \ll 1$. We may then approximate the cross-over time for the background equation of state, $\tilde{t}_{\rm cross}^{(w)}$, as the time when $6 \xi_\phi \delta_{\rm max}^2 (t_{\rm cross}^{(w)}) \simeq 1$, or
\beq
6 \xi_\phi \alpha^2 \simeq e^{3N/2} ,
\label{Ntildet2}
\eeq
upon using Eq.~(\ref{deltamaxN}). Comparing with Eq.~(\ref{Ntildet1}), we then find
\beq
 \tilde{t}_{\rm cross}^{(w)} \simeq \frac{ 24} {\sqrt{3}  } \, \xi_\phi .
\label{tcrossw}
\eeq
Hence we find the scaling $\tilde{t}_{\rm cross}^{(w)} \sim \xi_\phi$, akin to the large-$\xi_\phi$ scaling we had found for the cross-over among dominant contributions to the effective mass of the isocurvature perturbations, $\tilde{t}_{\rm cross}^{(\chi)}$.

Though the similarity in scaling between $\tilde{t}_{\rm cross}^{(w)}$ and $\tilde{t}_{\rm cross}^{(\chi)}$ is interesting, we note that $w_{\rm avg}$ is calculated to {\it background-order} only, and does not take into account the contributions to the total energy density from particles produced during preheating. Hence $w_{\rm avg}$, as we have calculated it numerically in Ref.~\cite{MultiPreheat1} and estimated its cross-over time here, is independent of the production of (say) $\chi$ particles. Rather, we may understand the similar scaling of $\tilde{t}_{\rm cross}^{(w)}$ and $\tilde{t}_{\rm cross}^{(\chi)}$ with $\xi_\phi$ by looking more closely at several relevant quantities.

The behavior of $w_{\rm avg}$, and hence of $\tilde{t}_{\rm cross}^{(w)}$, depends on the ratio $V (\phi^I) / \dot{\sigma}^2 $, while the behavior of $m_{\rm eff,\chi}^2$, and hence of $\tilde{t}_{\rm cross}^{(\chi)}$, depends on the ratio $m_{1,\chi}^2 / m_{2,\chi}^2 $. Within a single-field attractor along $\chi \sim 0$, we have, from Eqs.~(\ref{m2chifirst}) and (\ref{dotsigma1}),
\beq
m_{2,\chi}^2 = \frac{1}{2} {\cal R} \dot{\sigma}^2 ,
\label{m2app1}
\eeq
where ${\cal R}$ is the Ricci scalar for the field-space manifold. In the regime $\xi_I \gg 1$, $\delta^2 \ll 1$, and $\xi_\phi \delta^2 \sim {\cal O} (1)$, along $\chi \sim 0$, we have
\beq
{\cal R} \simeq \frac{12 \xi_\phi \xi_\chi}{ M_{\rm pl}^2 [1 + 6 \xi_\phi \delta^2 ]^2 } ,
\label{Riccilimit}
\eeq
upon using the expressions in Appendix A of Ref.~\cite{MultiPreheat1}. In that same regime, we find
\beq
m_{1,\chi}^2 = {\cal G}^{\chi K}\left( {\cal D}_\chi {\cal D}_K V \right) \simeq  \frac{ 6 \vert \Lambda_\phi \vert \, M_{\rm pl}^2}{\xi_\phi} \frac{ \delta^4}{(1 + 6 \xi_\phi \delta^2 ) } ,
\label{m1chiappendix}
\eeq
where $\Lambda_\phi \equiv \lambda_\phi \xi_\chi - g \xi_\phi$. Then in the regime of interest, we find 
\beq
\frac{ m_{1,\chi}^2 }{m_{2,\chi}^2} \sim \frac{ \vert \Lambda_\phi \vert \, M_{\rm pl}^4 }{\xi_\chi \xi_\phi^2  } \frac{ \delta^4}{\dot{\sigma}^2} (1 + 6 \xi_\phi \delta^2 ) ,
\label{m1m2ratioappendix}
\eeq
while for this same regime, we find 
\beq
\frac{V}{\dot{\sigma}^2} \sim \frac{ \lambda_\phi \, M_{\rm pl}^4}{\xi_\phi^2} \frac{\delta^4}{\dot{\sigma}^2 } .
\label{Vsigmaratio}
\eeq
For the case of weakly broken symmetries, with $\lambda_\phi \sim g$ and $\xi_\phi \sim \xi_\chi$, of the sort we examined throughout Sections \ref{sec:Adiabatic} and \ref{sec:Isocurvature}, we can expect $\vert \Lambda_\phi \vert / \xi_\chi \sim \lambda_\phi$. Then we find that the ratios in Eqs.~(\ref{m1m2ratioappendix}) and (\ref{Vsigmaratio}) scale quite similarly with parameters, especially around the critical transition period when $6 \xi_\phi \delta^2 \sim 1$. The cross-over time $\tilde{t}_{\rm cross}^{(\chi)}$ depends on the ratio in Eq.~(\ref{m1m2ratioappendix}), while the cross-over time $\tilde{t}_{\rm cross}^{(w)}$ depends on the ratio in Eq.~(\ref{Vsigmaratio}), and hence we may understand why these two distinct cross-over times scale in a comparable way with $\xi_\phi$.

\acknowledgements{It is a pleasure to thank Mustafa Amin,  Bruce Bassett,  Jolyon Bloomfield,  Peter Fisher,  Tom Giblin,  Alan Guth,  Mark Hertzberg, and Johanna Karouby for helpful discussions. We also thank the anonymous referee for helpful comments and suggestions. We would like to acknowledge support from the Center for Theoretical Physics at MIT. This work is supported by the U.S. Department of Energy under grant Contract Number DE-SC0012567. 
MPD and AP were also supported in part by MIT's Undergraduate Research Opportunities Program (UROP). CPW thanks the University of Washington College of Arts \& Sciences for financial support. She also gratefully acknowledges support from the MIT Dr. Martin Luther King, Jr. Visiting Professors and Scholars program and its director Edmund Bertschinger.
EIS gratefully acknowledges support from a Fortner Fellowship at the University of Illinois at Urbana-Champaign. }


\begin{thebibliography}{999}
\bibitem{SMfn} It is also possible that the electroweak symmetry was never thermally restored, in which case perturbative decays of the Higgs into Standard Model particles would be occurring from the onset of reheating.
\end{thebibliography}
\end{document}